\documentclass[aps,prd,preprint,superscriptaddress,amsmath,amssymb,floatfix]{revtex4-2}

\usepackage{float}
\usepackage{graphicx}
\usepackage[T1]{fontenc}
\usepackage[utf8]{inputenc}
\usepackage[american]{babel}
\usepackage{etoolbox}
\usepackage{lmodern}
\usepackage{microtype}
\usepackage{mathtools}
\usepackage{bm}
\usepackage{comment}
\usepackage{tikz}
\usetikzlibrary{arrows.meta}
\usepackage[colorlinks=true,linkcolor=blue,citecolor=blue,urlcolor=blue]{hyperref}

\definecolor{domainblue}{RGB}{70,105,150}
\definecolor{domaingreen}{RGB}{55,125,95}
\definecolor{cutoffred}{RGB}{170,75,70}

\allowdisplaybreaks[2]

\newcommand{\dd}{\mathrm{d}}
\newcommand{\I}{\mathrm{i}}
\newcommand{\cA}{\mathcal{A}}
\newcommand{\cB}{\mathcal{B}}
\newcommand{\cM}{\mathcal{M}}
\newcommand{\cR}{\mathcal{R}}

\newcommand{\Order}{\mathcal{O}}

\begin{document}

\title{Harrison transport of boundary counterterms: renormalized action
and quasilocal energy of Schwarzschild--Melvin}

\author{David Choque Quispe}
\email{dchoque@pucp.edu.pe}
\affiliation{Secci\'on F\'isica, Departamento de Ciencias,
Pontificia Universidad Cat\'olica del Per\'u,
Av. Universitaria 1801, San Miguel, Lima, Peru}
\author{Alessandro Huaman}
\email{huaman.alessandro@pucp.edu.pe}
\affiliation{Secci\'on F\'isica, Departamento de Ciencias,
Pontificia Universidad Cat\'olica del Per\'u,
Av. Universitaria 1801, San Miguel, Lima, Peru}

\author{Guillermo Ram\'{i}rez-Ulloa}
\email{gramirezu@unitru.edu.pe}
\affiliation{
 Departamento de F\'{i}sica, Universidad Nacional de Trujillo, Avenida Juan Pablo II S/N; Ciudad Universitaria, Trujillo, La Libertad, Per\'{u}
}%

\author{Gustavo Alonso Tapia Mantilla}
\email{gustavo.tapia@upn.pe}
\affiliation{
 Departamento de Ciencias, UPN - Universidad Privada del Norte, Trujillo, La Libertad, Perú}%

\begin{abstract}
Counterterm methods for asymptotically flat spacetimes do not extend
automatically to black holes immersed in external magnetic fields. We show
that the intrinsic counterterm $\sqrt{2\cR}$, with $\cR$ the scalar
curvature of the boundary, does not renormalize the Einstein--Maxwell
action of the Melvin and Schwarzschild--Melvin spacetimes: on complete
constant-radius boundaries it leaves a divergence $\beta\ell\ln(r/\ell)$,
where $\beta$ is the Euclidean period and $B_0=2/\ell$ is the magnetic
field on the axis, in contrast with an earlier statement in the
literature. The divergence originates in narrow regions around the
symmetry axis. We then prove that the Harrison transformation, which
generates both solutions from Minkowski and Schwarzschild, leaves the
boundary density $\sqrt{h}\,\sqrt{2(\cR-\mathcal F^2)}$ invariant when
the axial Killing vector is hypersurface orthogonal, where $h$ is the
induced metric and $\mathcal F$ the Maxwell field induced on the
boundary. The Harrison transformation therefore acts on the renormalized action
and not only on the solutions: it maps the counterterm $\sqrt{2\cR}$ of
the asymptotically flat seeds to $\sqrt{2(\cR-\mathcal F^2)}$, including
its coefficient, and the renormalized action of each magnetized solution
equals that of its seed at every cutoff, without background subtraction. Varying the
full action with respect to the induced metric and the tangential
potential, we obtain the Brown--York tensor and a boundary current. Both
vanish identically for Melvin, together with the action, at every cutoff.
For Schwarzschild--Melvin, with mass parameter $m$, the action tends to
$\beta m/2$ and the Killing-time charge, normalized on the axis,
tends to $m$, in agreement with the values obtained by background
subtraction. The local responses depend on the field even though the
charge does not: the boundary stress tensor is anisotropic, and the
boundary current, absent in Melvin, is proportional to $mB_0$ at leading
order. The first law holds at fixed $B_0$, and the response conjugate
to $B_0$ vanishes.
\end{abstract}

\maketitle

\newpage
\section{Introduction}
\label{sec:introduction}

The Melvin universe~\cite{Melvin:1964} and the Schwarzschild--Melvin
solution, which describes a black hole immersed in the external Melvin
field~\cite{Ernst:1976,Radu:2002,Ortaggio:2005}, are the simplest
Einstein--Maxwell geometries in which a magnetic field modifies the
asymptotic structure. These geometries are not asymptotically flat, and
defining their global quantities requires specifying both the boundary
prescription and the normalization of the time-translation generator.
To our knowledge, no counterterm prescription has been established for
asymptotically Melvin spacetimes. The Euclidean action of
Schwarzschild--Melvin has been obtained by subtracting the Melvin
background~\cite{Radu:2002}, and the mass of magnetized Kerr--Newman
black holes has been the subject of different
proposals~\cite{GibbonsPangPope:2014,BoothEtAl:2015,AstorinoEtAl:2016}.
The question has received renewed attention for black holes immersed in
external Bertotti--Robinson and Bonnor--Melvin fields~\cite{Bonnor:1954,Melvin:1964,PodolskyOvcharenko:2025,Astorino:2025}, whose thermodynamics
has so far been formulated with Komar integrals, the
Christodoulou--Ruffini mass relation, or background
subtraction~\cite{KubiznakOvcharenkoPodolsky:2026,HuCaiWang:2026,YanEtAl:2026}.

The action with the Gibbons--Hawking--York term and the
Brown--York procedure relate the boundary data
to the gravitational responses and charges
\cite{GibbonsHawking,BrownYork,HawkingHorowitz}.
Two routes are available to remove its divergences. Background
subtraction compares the solution with a reference spacetime into
which its boundary must be embedded, or with which it must be matched
asymptotically~\cite{GibbonsHawking,BrownYork,HawkingHorowitz}. The
subtracted term then depends on the reference and on the matching, not
only on the boundary data of the solution, and a suitable embedding
need not exist for general boundaries~\cite{Mann:1999,KrausLarsenSiebelink}.
For Schwarzschild--Melvin the reference is the Melvin background, and
Radu already noted that the procedure is in general ill defined and
can lead to ambiguities~\cite{Radu:2002}. Local counterterms replace
the choice of a reference by the specification of the boundary data
held fixed. Two tasks must then be distinguished: controlling the
value of the action as the regulator is removed, and deriving the
boundary responses from its complete first variation with respect to
those data. The cancellation of divergences on a solution does not
determine the boundary tensor or the charges. The question guiding this
work is how to carry out both tasks for Melvin and
Schwarzschild--Melvin by means of a geometrically justified
intrinsic counterterm and without background subtraction.

A natural starting point is the counterterm $\sqrt{2\cR[h_E]}$ used for
boundaries of topology $S^1\times S^2$ in asymptotically flat
spacetimes, where $h_E$ is the Euclidean induced metric and $\cR[h_E]$
its three-dimensional scalar curvature
\cite{Mann:1999,KrausLarsenSiebelink,AstefaneseiMannStelea:2007}.
We evaluate it on the complete boundaries $r=\mathrm{const.}$ of the two
geometries, whose spatial sections are topological spheres, keeping
fixed the magnetic field on the axis, $B_0=2/\ell$. With $\beta$ the
Euclidean period and $\kappa=8\pi$ in units $G_N=c=1$, the sum of the
bulk, Gibbons--Hawking--York, and counterterm contributions is
\begin{equation}
I^E_{\mathrm{bulk}}+I^E_{\mathrm{GH}}
+\frac1\kappa\int_\Sigma\dd^3x\,\sqrt{h_E}\sqrt{2\cR[h_E]}
=\beta\ell\ln\frac r\ell+\Order(1),
\qquad r\to\infty,
\label{eq:intro-log}
\end{equation}
for both solutions. The divergence originates in narrow regions around
the symmetry axis (Fig.~\ref{fig:polar-annulus}), where the expansion at fixed polar angle fails; the
densities are regular there at finite radius. This result is in contrast with the
statement of Ref.~\cite{Radu:2002} that the same functional yields a
finite action. The leading term of the fixed-angle expansion does
reproduce a finite value; the divergence comes from the polar regions
where that expansion fails, as detailed in
Sec.~\ref{sec:geometric-counterterm-limit}. The value $\beta m/2$
obtained in~\cite{Radu:2002} through subtraction of the Melvin
background is unaffected.

The alternative we develop starts from the Harrison
transformation, which generates Melvin and Schwarzschild--Melvin from
Minkowski and Schwarzschild
\cite{Harrison:1968,Ernst:1976,Ortaggio:2005}.
We restrict ourselves to the hypersurface-orthogonal axial sector, in
which the axial Killing vector is hypersurface orthogonal and the
potential is aligned with it. Keeping track of the boundary terms in
the axial reduction, we establish an intrinsic identity that transports
the density of the seed counterterm, including the factor $\sqrt{h_E}$:
\begin{equation}
\sqrt{h_E}\sqrt{2\bigl(\cR[h_E]-\mathcal F^2\bigr)}
=\sqrt{\hat h_E}\sqrt{2\cR[\hat h_E]} .
\label{eq:intro-transport}
\end{equation}
Hats denote the neutral seed, $\mathcal F=\dd(A|_\Sigma)$ is the Maxwell
field induced on the boundary $\Sigma$, and $\mathcal F^2$ is contracted
with $h_E$; it is not the restriction of the four-dimensional invariant
$F^2$. The geometric functional of the seed thus leads to the
counterterm
\begin{equation}
I^E_{\mathrm{ct}}
=\frac1\kappa\int_\Sigma\dd^3x\,
\sqrt{h_E}\sqrt{2\bigl(\cR[h_E]-\mathcal F^2\bigr)} .
\label{eq:intro-counterterm}
\end{equation}
The electromagnetic term and its coefficient follow
from the transport considered, not from a fit to the final value
of the action or of the mass. Together with the invariance of the bulk and
Gibbons--Hawking--York terms, the identity implies that the
renormalized action of each magnetized solution equals that of its seed
at every cutoff, Eq.~\eqref{eq:harrison-action-invariance}. This
accounts for the independence of the action of the external field noted
in~\cite{Radu:2002,Ortaggio:2005}.

The transport identity does not replace the variation with respect
to the independent induced metric and tangential potential.
We vary the complete functional before evaluating it on the solutions,
including the metric and electromagnetic contributions
of the counterterm, and obtain the Brown--York tensor and the
boundary current conjugate to the tangential potential. Since the
counterterm is intrinsic, Dirichlet stationarity at finite cutoff is
preserved; the responses, by contrast, are not fixed by the transport
identity, and they carry the content of the construction that is
independent of the value of the action.

The contributions of this work are the following. (i) We show that
the counterterm $\sqrt{2\cR}$ does not renormalize the action of Melvin
and Schwarzschild--Melvin on complete constant-radius boundaries: it
leaves the logarithmic divergence of Eq.~\eqref{eq:intro-log}.
(ii) We show that the Harrison transformation acts on the renormalized
action and not only on the solutions. It leaves the boundary density of
Eq.~\eqref{eq:intro-transport} invariant in the hypersurface-orthogonal
axial sector, so that the counterterm $\sqrt{2\cR}$ of the
asymptotically flat seeds is mapped to
Eq.~\eqref{eq:intro-counterterm}, including its coefficient, and the
renormalized action of each magnetized solution equals that of its
seed. The new counterterm is thus traced to a symmetry of the
Einstein--Maxwell action, not fitted to the magnetized solutions.
(iii) We specify the boundary data held fixed, namely the induced
metric and the tangential potential in a regular gauge, and derive from
the complete variation the Brown--York tensor and the boundary current.
For Melvin they vanish, together with the action, at every cutoff,
which fixes the zero of the quasilocal energy in this sector. For
Schwarzschild--Melvin the action tends to
$I^{E,\infty}_{\mathrm{ren}}=\beta m/2$ and the Killing-time charge
tends to $Q_\infty[\partial_t]=m$, with $\beta=8\pi m$; the charge
includes the lapse and is distinct from the proper energy of the static
observers. The local responses depend on the field even though the
charge does not: the anisotropy of the boundary stress tensor and the
boundary current require both the black hole and the external field,
and at leading order the current is proportional to $mB_0$. Action and
charge satisfy the first law at fixed $B_0$ without background
subtraction and reproduce the known
values~\cite{Radu:2002,Ortaggio:2005}; the response conjugate to $B_0$
vanishes, as does the magnetic moment of an uncharged magnetized black
hole in Ref.~\cite{GibbonsPangPope:2014}. The invariance of the density
holds in the whole hypersurface-orthogonal axial sector, as we check
explicitly for seeds with nonvanishing potential in
Appendix~\ref{app:harrison-charged-seeds}; rotating solutions such as
Kerr--Melvin lie outside it. The explicit evaluations are restricted to
the two families and to the complete spherical regulator considered,
and the uniqueness of the counterterm is not established.

Section~\ref{sec:melvin-universe} fixes the theory, the solutions,
and the boundary data.
Section~\ref{sec:geometric-counterterm-limit} examines the
geometric prescription, and
Sec.~\ref{sec:harrison-construction} develops the Harrison
transport. Sections~\ref{sec:variational-principle},
\ref{sec:euclidean-action}, and \ref{sec:brown-york-energy}
present the first variation, the Euclidean action, and the
quasilocal responses, respectively.
Section~\ref{subsec:schmel-thermodynamic-consistency}
collects the thermodynamic checks, and
Sec.~\ref{app:counterterm-checks} discusses the results
and their limits. The appendices contain the polar integrals,
the details of the axial reduction, the integrals of the action,
and the local components of the responses.

\section{Theory, solutions, and spherical boundaries}
\label{sec:melvin-universe}

We fix the Einstein--Maxwell conventions and the geometric data
of the boundaries that we will use. The distinction between bulk
fields and induced fields will be essential both for examining the
purely geometric counterterm and for constructing its extension
by means of the Harrison transformation.

\subsection{Einstein--Maxwell action and conventions}

We work with signature $(-+++)$ and units $G_N=c=1$, so
that $\kappa=8\pi$. The electromagnetic field is $F=\dd A$ and
$F^2=F_{\mu\nu}F^{\mu\nu}$, with indices raised with the
bulk metric $g_{\mu\nu}$. Greek indices refer
to the bulk and $a,b$ to the three-dimensional boundary $\Sigma$;
$\nabla_\mu$ and $D_a$ denote the respective connections.

The Lorentzian action, including the
Gibbons--Hawking--York term~\cite{GibbonsHawking}, is
\begin{equation}
  I_{\mathrm{bulk+GH}}
  =\frac{1}{2\kappa}
   \int_{\cM}\dd^4x\,\sqrt{-g}\left(R-F^2\right)
   +\frac{1}{\kappa}
   \int_{\partial\cM}\dd^3x\,\sqrt{-h}\,K .
  \label{eq:lorentzian-bulk-GH}
\end{equation}
On the radial boundaries we use the outward unit spacelike
normal and the convention
\begin{equation}
  n^\mu n_\mu=1,\qquad
  h_{\mu\nu}=g_{\mu\nu}-n_\mu n_\nu,\qquad
  K_{ab}=h_a{}^\mu h_b{}^\nu\nabla_\mu n_\nu .
  \label{eq:extrinsic-curvature-convention}
\end{equation}
The corresponding equations of motion are
\begin{equation}
  G_{\mu\nu}
  =2\left(
    F_{\mu\rho}F_\nu{}^\rho-\frac14g_{\mu\nu}F^2
    \right),
  \qquad
  \nabla_\mu F^{\mu\nu}=0 .
\end{equation}

In Euclidean signature we write $g_E$ and $h_E$ for the metrics,
and for their determinants when they appear under a square root.
We reserve $\cR$ for the scalar curvature of the three-dimensional
boundary. The induced field is
$\mathcal F=\dd(A|_\Sigma)$ and its invariant is
$\mathcal F^2=\mathcal F_{ab}\mathcal F^{ab}$, with indices
raised with the induced metric. This scalar must not be
identified with the restriction to $\Sigma$ of the
four-dimensional invariant $F^2$.

\subsection{Melvin and Schwarzschild--Melvin geometries}

We consider $\ell>0$ and an axial angle of period $2\pi$.
In spherical coordinates, the Melvin solution~\cite{Melvin:1964} reads
\begin{equation}
\begin{aligned}
  \dd s^2&=\Lambda^2
       \left(-\dd t^2+\dd r^2+r^2\dd\theta^2\right)
       +\frac{r^2\sin^2\theta}{\Lambda^2}\dd\varphi^2,\\
  A&=\frac{r^2\sin^2\theta}{\ell\Lambda}\dd\varphi,
  \qquad
  \Lambda=1+\frac{r^2\sin^2\theta}{\ell^2}.
\end{aligned}
  \label{eq:harrison-Melvin-spherical}
\end{equation}
The electromagnetic components that we will use are
\begin{equation}
  F_{r\varphi}=\frac{2r\sin^2\theta}{\ell\Lambda^2},
  \qquad
  F_{\theta\varphi}
       =\frac{2r^2\sin\theta\cos\theta}{\ell\Lambda^2}.
\end{equation}
The field strength on the axis is $B_0=2/\ell$.

Schwarzschild--Melvin describes a black hole immersed in
the external Melvin field~\cite{Radu:2002,Ortaggio:2005}.
Its metric is
\begin{equation}
  \dd s^2=\Lambda^2\left(
       -f(r)\dd t^2+\frac{\dd r^2}{f(r)}
       +r^2\dd\theta^2\right)
       +\frac{r^2\sin^2\theta}{\Lambda^2}\dd\varphi^2,
  \label{eq:schmel-sph-metric}
\end{equation}
where
\begin{equation}
  f(r)=1-\frac{2m}{r},
  \qquad
  \Lambda=1+\frac{r^2\sin^2\theta}{\ell^2}.
\end{equation}
The potential and the components of $F$ retain the forms
given above. For $m>0$, the horizon is located at $r=2m$.
The $m=0$ limit of the metric recovers Melvin, whereas
$\ell\to\infty$ at fixed radius recovers Schwarzschild.

The term spherical refers to the foliation and to the
topology of its spatial sections: for nonzero field,
the geometry is axially symmetric and the surfaces
$r=\mathrm{const.}$ are not round spheres. The derivation of
both solutions from Minkowski and Schwarzschild
by means of the Harrison map is developed in
Sec.~\ref{subsec:harrison-generation}
\cite{Harrison:1968,Ernst:1976,Ortaggio:2005}.

We keep the time normalization of these metrics.
In Melvin, $\xi=\partial_t$ corresponds to unit proper
time on the axis, where $\Lambda=1$. In Schwarzschild--Melvin,
the inherited normalization satisfies $-\xi^2\to1$ on the
asymptotic axis. This choice does not make the generator
of unit norm over the whole boundary.

\subsection{Regulator and induced data}

The outer Lorentzian boundaries have topology
$\mathbb R_t\times S^2$ and are located at constant radius:
$r>0$ for Melvin and $r>2m$ for Schwarzschild--Melvin.
To describe the Euclidean regions we use
$t=-i\tau$ and $I^E=-iI$. From now on we distinguish
the cutoff radius $r$ from the radial integration
coordinate $\bar r$, and write
\begin{equation*}
  \bar\Lambda=1+\frac{\bar r^2\sin^2\theta}{\ell^2},
  \qquad
  0\leq\theta\leq\pi,\qquad
  \tau\sim\tau+\beta,\qquad
  \varphi\sim\varphi+2\pi .
\end{equation*}
The unbarred function $\Lambda$ denotes its value at $\bar r=r$.

For Melvin, $0\leq\bar r\leq r$ and the period $\beta$ is
arbitrary. The region has topology
$\cM_r\simeq S^1_\beta\times B^3$ and a single smooth,
closed boundary,
\begin{equation}
  \partial\cM_r=\Sigma:\bar r=r,
  \qquad
  \Sigma\simeq S^1_\beta\times S^2 .
  \label{eq:melvin-sph-boundary}
\end{equation}
The origin and the poles are regular degenerations of the
coordinates, not additional components of $\partial\cM_r$.

For Schwarzschild--Melvin we regulate the exterior by
$2m\leq\bar r\leq r$, with $r>2m$. The outer boundary
likewise has topology $S^1_\beta\times S^2$.
In the regular Euclidean solution, the horizon is not an
inner boundary: the Euclidean circle shrinks there and
regularity fixes $\beta=8\pi m$, as shown by
Eq.~\eqref{eq:schmel-sph-bolt-regularity}. No
Gibbons--Hawking--York term is introduced on that set
as if it were an inner boundary
\cite{GibbonsHawking,HawkingHorowitz}.

\paragraph{Induced metric and outward normal.}

In Melvin, the Euclidean data at the cutoff are
\begin{equation}
\begin{aligned}
  \dd s^2_{\Sigma,E}
  &=\Lambda^2\dd\tau^2+\Lambda^2r^2\dd\theta^2
    +\frac{r^2\sin^2\theta}{\Lambda^2}\dd\varphi^2,
  \qquad \sqrt{h_E}=\Lambda r^2\sin\theta,\\
  n_\mu\dd x^\mu
  &=\left.\bar\Lambda\,\dd\bar r\right|_\Sigma,
  \qquad
  n^\mu\partial_\mu
  =\left.\bar\Lambda^{-1}\partial_{\bar r}\right|_\Sigma .
\end{aligned}
  \label{eq:melvin-sph-induced-normal}
\end{equation}
With the convention fixed above, the trace of the
extrinsic curvature and its density are
\begin{equation}
\begin{aligned}
  K&=\left.
       \frac{1}{\bar\Lambda^2\bar r^2}
       \partial_{\bar r}(\bar\Lambda\bar r^2)
       \right|_{\bar r=r}
     =\frac{2+4r^2\sin^2\theta/\ell^2}{r\Lambda^2},\\
  \sqrt{h_E}\,K
     &=2r\sin\theta\left(2-\frac1\Lambda\right).
\end{aligned}
  \label{eq:melvin-sph-K}
\end{equation}
The radial derivative is taken before restricting to the cutoff.

In Schwarzschild--Melvin, the outward normal is
$n=(\sqrt f/\Lambda)\partial_{\bar r}$, evaluated at
$\bar r=r$, and the corresponding data are
\begin{equation}
\begin{aligned}
  \dd s_{\Sigma,E}^2
    &=\Lambda^2f(r)\dd\tau^2+\Lambda^2r^2\dd\theta^2
      +\frac{r^2\sin^2\theta}{\Lambda^2}\dd\varphi^2,\\
  \sqrt{h_E}
    &=\Lambda\sqrt f\,r^2\sin\theta,\\
  K&=\frac{\sqrt f}{\Lambda}
     \left[\frac4r-\frac{2}{r\Lambda}
                  +\frac{m}{r(r-2m)}\right].
\end{aligned}
  \label{eq:schmel-sph-boundary-geometry}
\end{equation}
Here $f=f(r)>0$ is constant with respect to the tangential
coordinates of each boundary.

\paragraph{Induced field and intrinsic invariants.}

The tangential restriction can be written as
$\mathcal F_{ab}=h_a{}^\mu h_b{}^\nu F_{\mu\nu}$.
It contains $\mathcal F_{\theta\varphi}
=2r^2\sin\theta\cos\theta/(\ell\Lambda^2)$, but not
$F_{r\varphi}$. For Melvin,
\begin{equation}
  \mathcal F^2
     =\frac{8\cos^2\theta}{\ell^2\Lambda^4}.
  \label{eq:melvin-sph-pullback-F2}
\end{equation}
The curvature of the three-dimensional boundary satisfies
\begin{equation}
  \cR=\frac{2}{r^2\Lambda^2}
        +\frac{8\cos^2\theta}{\ell^2\Lambda^4},
  \qquad
  \cR-\mathcal F^2=\frac{2}{r^2\Lambda^2}>0 .
  \label{eq:melvin-sph-intrinsic-invariants}
\end{equation}

In Schwarzschild--Melvin, the local change
$\widetilde\tau=\sqrt{f(r)}\,\tau$ maps the induced
metric to that of Melvin at the same $r$ and $\ell$.
The global period becomes $\beta\sqrt f$, but the
intrinsic scalars remain the same:
\begin{equation}
  \mathcal F^2=\frac{8\cos^2\theta}{\ell^2\Lambda^4},
  \qquad
  \cR=\frac{2}{r^2\Lambda^2}
        +\frac{8\cos^2\theta}{\ell^2\Lambda^4}.
  \label{eq:schmel-sph-intrinsic-invariants}
\end{equation}
This local equality does not identify the Euclidean boundaries
at equal global period. In both expressions, $\cR$ is the
three-dimensional curvature of the boundary, not the curvature
of the bulk nor that of its spatial section. Likewise,
$\mathcal F^2$ excludes the components with one normal
index that contribute to $F^2$.

\paragraph{Axial regularity.}

At both poles, $\Lambda=1+\Order(\sin^2\theta)$ and the
azimuthal period $2\pi$ avoids conical defects.
The regularity of the potential can be made explicit in
the local coordinates
$u=\sin\theta\cos\varphi$ and $v=\sin\theta\sin\varphi$:
\begin{equation}
  A\big|_\Sigma
    =\frac{r^2}{\ell[1+r^2(u^2+v^2)/\ell^2]}
       (u\,\dd v-v\,\dd u).
  \label{eq:melvin-sph-polar-potential}
\end{equation}
This expression is smooth at $u=v=0$ and applies to both
families. The poles are not additional edges of the
boundary, so the subsequent analysis will retain
the complete sphere.

\section{Limitation of the purely geometric counterterm}
\label{sec:geometric-counterterm-limit}

We examine the purely geometric prescription on the complete
spherical boundaries of Melvin and Schwarzschild--Melvin. We distinguish
three issues: the regularity of the integral at finite cutoff,
the nonuniformity of its radial expansion, and the behavior
of the action after integration. The last issue requires
control of the complete integral; it does not follow solely from
the expansion at fixed angle.

\subsection{Prescription and regularity at finite cutoff}

The three-dimensional intrinsic functional is
\begin{equation}
  I^E_{\mathrm{ct,g}}
  =\frac1\kappa\int_\Sigma\dd^3x\,\sqrt{h_E}\sqrt{2\cR}
  \label{eq:melvin-sph-pure-counterterm}
\end{equation}
Its antecedents lie in the treatment of asymptotically
flat boundaries
\cite{Mann:1999,KrausLarsenSiebelink,AstefaneseiMannStelea:2007}.

The evaluation is carried out over the whole sphere, with periods
$\beta$ and $2\pi$, at fixed $\ell>0$ and without additional
background subtraction. Using the data of Sec.~\ref{sec:melvin-universe},
the Melvin density is
\begin{equation}
  \sqrt{h_E}\sqrt{2\cR}
  =2r\sin\theta\left[
      1+\frac{4r^2\cos^2\theta}{\ell^2\Lambda^2}\right]^{1/2}.
  \label{eq:melvin-sph-pure-density}
\end{equation}
We define $x=r/\ell$ and denote by $\vartheta$ the angular
distance to the corresponding pole: $\vartheta=\theta$ in the north
and $\vartheta=\pi-\theta$ in the south. At fixed radius,
\begin{equation}
  \sqrt{h_E}\sqrt{2\cR}
  =2r\sqrt{1+4x^2}\,\vartheta+\Order_r(\vartheta^3),
  \qquad \vartheta\to0.
  \label{eq:melvin-sph-polar-regularity}
\end{equation}
The subscript of $\Order_r$ indicates that the polar limit is taken
at fixed $r$. The density vanishes linearly and its integral
is regular at both poles.

In Schwarzschild--Melvin, $f(r)=1-2m/r>0$ is constant on
each outer cutoff $r>2m$. It does not modify the intrinsic scalars
of Eq.~\eqref{eq:schmel-sph-intrinsic-invariants}, but it multiplies
$\sqrt{h_E}$ by $\sqrt f$. Thus,
\begin{equation}
  w_g(r,\theta)\equiv\sqrt{h_E}\sqrt{2\cR}
   =2r\sqrt f\,\sin\theta\sqrt{1+q_x(\theta)},
  \qquad
  q_x=\frac{4x^2\cos^2\theta}{(1+x^2\sin^2\theta)^2},
  \quad x=\frac r\ell.
  \label{eq:schmel-sph-pure-density}
\end{equation}
and its polar behavior is
\begin{equation}
  w_g=2r\sqrt f\,\sqrt{1+4x^2}\,\vartheta
           +\Order_r(\vartheta^3).
  \label{eq:schmel-sph-pure-polar-regularity}
\end{equation}
Here too the poles are integrable and $\cR>0$ guarantees
that the square root is real. The problem we examine is neither a
physical singularity on the axis nor a divergence of the
integral at finite radius.

\subsection{Nonuniform expansion and polar contribution}

In Melvin, the ratio between the two contributions to the intrinsic
curvature exhibits the nonuniformity:
\begin{equation}
  q_x(\theta)
   =\frac{8\cos^{2}\theta/\ell^2\Lambda^{4}}{2/(r^2\Lambda^2)}
   =\frac{4x^2\cos^2\theta}{(1+x^2\sin^2\theta)^2}.
  \label{eq:melvin-sph-polar-ratio}
\end{equation}
For every fixed $0<\theta<\pi$, $q_x\to0$ as $x\to\infty$;
however, $q_x(0)=q_x(\pi)=4x^2$. At the poles,
$\cR=2/r^2+8/\ell^2$, so the electromagnetic term
cannot be neglected there. The expansion at fixed angle gives
\begin{equation}
  \sqrt{h_E}\sqrt{2\cR}
   =2r\sin\theta+
     \frac{4\ell^2}{r}\frac{\cos^2\theta}{\sin^3\theta}
     +\Order_\theta\!\left(\frac{\ell^4}{r^3}\right).
  \label{eq:melvin-sph-pure-fixed-angle}
\end{equation}
The first correction behaves as $\vartheta^{-3}$ and is not
integrable at the endpoints. Therefore, this expansion cannot
be integrated term by term up to the poles. The
divergence of its coefficients is not the behavior of the
exact density~\eqref{eq:melvin-sph-polar-regularity}.
The corresponding expression for Schwarzschild--Melvin,
Eq.~\eqref{eq:schmel-sph-pure-fixed-angle}, retains the same
angular problem.

There are two different scales. Approximating
$\Lambda\simeq x^2\vartheta^2$ requires
$\vartheta\gg x^{-1}$, whereas expanding
$\sqrt{1+q_x}$ requires $q_x\ll1$, that is,
$\vartheta\gg x^{-1/2}$. Between the two scales lies the annulus
$x^{-1}\ll\vartheta\ll x^{-1/2}$.
To isolate its contribution, we subtract exactly the density
$2r\sin\theta$, whose integral contributes $\beta r$:
\begin{equation}
\begin{aligned}
  g_x(\theta)&=x\sin\theta\,[\sqrt{1+q_x(\theta)}-1],\\
  I^E_{\mathrm{ct,g}}-\beta r
    &=\frac{\beta\ell}{2}\int_0^\pi g_x(\theta)\,\dd\theta
      =\beta\ell D(x),\qquad
  D(x)=\int_0^{\pi/2}g_x(\theta)\,\dd\theta.
\end{aligned}
  \label{eq:melvin-sph-polar-excess}
\end{equation}
In that annulus, $q_x\gg1$ and $g_x\sim2/\vartheta$.
The integral accumulates
$\int_{x^{-1}}^{x^{-1/2}}2\,\dd\vartheta/\vartheta=\ln x$;
each pole contributes $(\beta\ell/2)\ln x$ at logarithmic order,
as illustrated in Fig.~\ref{fig:polar-annulus}.
The control of the ends of the annuli is kept in
Appendix~\ref{app:prd-esfera-pura}, through
Eqs.~\eqref{eq:melvin-sph-polar-core-outer} and
\eqref{eq:melvin-sph-polar-annulus}. The exact evaluation and
the bound on the remainder will be used next to determine
the integrated action.
A fixed angular excision changes the integral under study:
for $0<\epsilon<\pi/2$,
$\int_\epsilon^{\pi-\epsilon}g_x\,\dd\theta
=\Order_\epsilon(x^{-1})$, whereas the complete integral
grows as $2\ln x$. Excluding those caps before taking
$r\to\infty$ does not compute the limit of the complete sphere.

\begin{figure}[htbp]
\centering
\begin{tikzpicture}[>=Stealth,font=\small]
\begin{scope}
\fill[domainblue!10] (-0.3,-2.9) rectangle (0.3,2.9);
\draw[domainblue!60,dashed] (-0.3,-2.9) -- (-0.3,2.9) (0.3,-2.9) -- (0.3,2.9);
\draw[->] (-2.8,0) -- (2.9,0) node[right] {$\rho$};
\draw[->] (0,-2.9) -- (0,3.2) node[above] {$z$};
\draw[cutoffred,thick] (0,0) circle (2.2);
\draw[domaingreen,line width=3pt] (82:2.2) arc[start angle=82,end angle=98,radius=2.2];
\draw[domaingreen,line width=3pt] (262:2.2) arc[start angle=262,end angle=278,radius=2.2];
\draw[domainblue,line width=3pt] (66:2.2) arc[start angle=66,end angle=82,radius=2.2];
\draw[domainblue,line width=3pt] (98:2.2) arc[start angle=98,end angle=114,radius=2.2];
\draw[domainblue,line width=3pt] (246:2.2) arc[start angle=246,end angle=262,radius=2.2];
\draw[domainblue,line width=3pt] (278:2.2) arc[start angle=278,end angle=294,radius=2.2];
\draw[gray,dashed] (0,0) -- (66:2.2);
\draw[gray,->] (90:1.1) arc[start angle=90,end angle=66,radius=1.1];
\node[gray] at (76:1.4) {$\vartheta$};
\node[domainblue,anchor=west] at (60:2.35) {annulus};
\node[domaingreen,anchor=east] at (101:2.4) {core};
\node[cutoffred,anchor=west] at (-40:2.3) {$r=\mathrm{const.}$};
\node[domainblue!70!black,anchor=north] at (0,-2.95) {$\rho\lesssim\ell$};
\node at (-2.6,3.0) {(a)};
\end{scope}
\begin{scope}[xshift=4.2cm,yshift=-1.5cm]
\draw[->] (0,0) -- (5.7,0);
\draw[->] (0,0) -- (0,3.3) node[above] {$\vartheta\,g_x(\vartheta)$};
\foreach \u/\lab in {0/-6,1.8/-4,3.6/-2,5.4/0} {\draw (\u,0) -- (\u,-0.08) node[below] {$\lab$};}
\foreach \y/\lab in {1.25/1,2.5/2} {\draw (0,\y) -- (-0.08,\y) node[left] {$\lab$};}
\node at (2.7,-0.75) {$\log_{10}\vartheta$};
\draw[gray,dotted] (0,2.5) -- (5.4,2.5);
\draw[gray,dotted] (0.9,0) -- (0.9,2.55);
\draw[gray,dotted] (3.15,0) -- (3.15,2.55);
\node[font=\scriptsize] at (0.9,2.8) {$x^{-1}$};
\node[font=\scriptsize] at (3.15,2.8) {$x^{-1/2}$};
\draw[domainblue,thick] plot[smooth] coordinates {(0.000,0.025) (0.090,0.039) (0.180,0.061) (0.270,0.096) (0.360,0.148) (0.450,0.227) (0.540,0.342) (0.630,0.502) (0.720,0.712) (0.810,0.967) (0.900,1.250) (0.990,1.533) (1.080,1.788) (1.170,1.998) (1.260,2.158) (1.350,2.273) (1.440,2.351) (1.530,2.404) (1.620,2.438) (1.710,2.460) (1.800,2.474) (1.890,2.482) (1.980,2.487) (2.070,2.489) (2.160,2.488) (2.250,2.485) (2.340,2.479) (2.430,2.468) (2.520,2.450) (2.610,2.422) (2.700,2.378) (2.790,2.310) (2.880,2.206) (2.970,2.051) (3.060,1.833) (3.150,1.545) (3.240,1.209) (3.330,0.874) (3.420,0.593) (3.510,0.387) (3.600,0.248) (3.690,0.157) (3.780,0.099) (3.870,0.063) (3.960,0.040) (4.050,0.025) (4.140,0.016) (4.230,0.010) (4.320,0.006) (4.410,0.004) (4.500,0.002) (4.590,0.002) (4.680,0.001) (4.770,0.001) (4.860,0.000) (4.950,0.000) (5.040,0.000) (5.130,0.000) (5.220,0.000) (5.310,0.000) (5.400,0.000)};
\draw[domaingreen,thick,dashed] plot[smooth] coordinates {(0.000,0.000) (0.090,0.000) (0.180,0.000) (0.270,0.000) (0.360,0.000) (0.450,0.000) (0.540,0.000) (0.630,0.000) (0.720,0.000) (0.810,0.000) (0.900,0.000) (0.990,0.000) (1.080,0.001) (1.170,0.001) (1.260,0.002) (1.350,0.002) (1.440,0.004) (1.530,0.006) (1.620,0.010) (1.710,0.016) (1.800,0.025) (1.890,0.039) (1.980,0.061) (2.070,0.096) (2.160,0.148) (2.250,0.227) (2.340,0.342) (2.430,0.502) (2.520,0.711) (2.610,0.966) (2.700,1.249) (2.790,1.531) (2.880,1.785) (2.970,1.993) (3.060,2.150) (3.150,2.260) (3.240,2.332) (3.330,2.373) (3.420,2.389) (3.510,2.383) (3.600,2.353) (3.690,2.294) (3.780,2.196) (3.870,2.045) (3.960,1.829) (4.050,1.542) (4.140,1.207) (4.230,0.872) (4.320,0.591) (4.410,0.385) (4.500,0.246) (4.590,0.156) (4.680,0.098) (4.770,0.062) (4.860,0.038) (4.950,0.024) (5.040,0.015) (5.130,0.009) (5.220,0.005) (5.310,0.003) (5.400,0.001)};
\draw[domainblue,thick] (4.0,3.05) -- (4.5,3.05) node[right,black,font=\scriptsize] {$x=10^5$};
\draw[domaingreen,thick,dashed] (4.0,2.75) -- (4.5,2.75) node[right,black,font=\scriptsize] {$x=10^3$};
\node at (-1.1,3.85) {(b)};
\end{scope}
\end{tikzpicture}
\caption{Localization of the logarithmic divergence left by
$\sqrt{2\cR}$. (a) Meridional section of the cutoff $r=\mathrm{const.}$
(red) in the $(\rho,z)$ plane, with the angular regions exaggerated.
Near each pole, the core $\vartheta\lesssim\ell/r$ (green) lies inside
the flux tube $\rho\lesssim\ell$ (shaded), and the annulus
$\ell/r\ll\vartheta\ll\sqrt{\ell/r}$ (blue) is the region where
$q_x\gg1$ although $\Lambda\simeq x^2\vartheta^2$.
(b) Integrand of $D(x)$, Eq.~\eqref{eq:melvin-sph-polar-excess}, per
unit $\ln\vartheta$ near one pole, for $x=r/\ell=10^3$ (dashed) and
$10^5$ (solid). The plateau $\vartheta g_x\simeq2$ between
$\vartheta=x^{-1}$ and $x^{-1/2}$, marked for $x=10^5$, has a width of
$\frac12\ln x$ in $\ln\vartheta$ and produces the term $\ln x$ of
Eq.~\eqref{eq:schmel-sph-polar-asymptotic}.}
\label{fig:polar-annulus}
\end{figure}
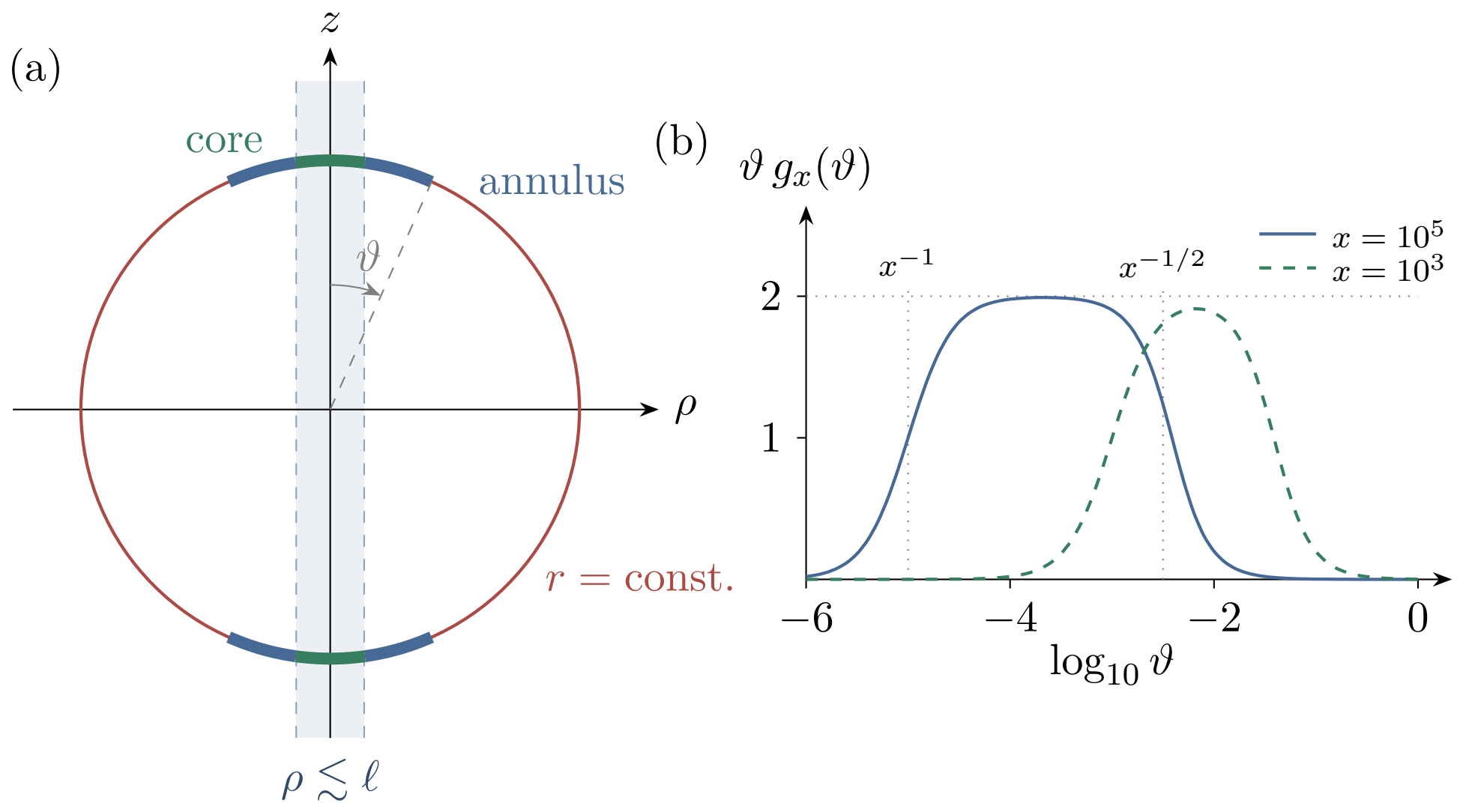
\subsection{Integrated divergence and scope of the comparison}
The exact change of variable of
Appendix~\ref{app:prd-esfera-pura} retains the polar regions.
The explicit antiderivative and the bound~\eqref{eq:polar-error-bound}
establish the result~\eqref{eq:polar-D-asymptotic}, which
we write here as
\begin{equation}
  D(x)=\ln(4x)-1+\Order\!\left(\frac{\ln x}{x}\right).
  \label{eq:schmel-sph-polar-asymptotic}
\end{equation}
To obtain the total action we use the exact sums of the
bulk and Gibbons--Hawking--York terms,
Eqs.~\eqref{eq:melvin-sph-bare-exact} and
\eqref{eq:schmel-sph-bare-exact}; their evaluations are presented
separately. In Melvin, the former is $-\beta r$, so that
\begin{equation}
\begin{aligned}
  I^E_{\mathrm{bulk}}+I^E_{\mathrm{GH}}+I^E_{\mathrm{ct,g}}
  &=\beta\ell\left[\ln\frac{4r}{\ell}-1\right] \\
  &\quad+\Order\!\left(
       \frac{\beta\ell^2}{r}\ln\frac{r}{\ell}\right).
\end{aligned}
  \label{eq:melvin-sph-pure-log}
\end{equation}
The counterterm cancels the linear divergence but leaves a
logarithmic divergence. This conclusion follows from the complete
integral with its remainder under control, not merely from the
nonuniformity of the angular expansion.

In Schwarzschild--Melvin, the same function $D(x)$ determines
the excess:
\begin{equation}
\begin{aligned}
  g_x(\theta)&=x\sin\theta\,[\sqrt{1+q_x(\theta)}-1],\\
  I^E_{\mathrm{ct,g}}-\beta r\sqrt f
    &=\frac{\beta\ell\sqrt f}{2}\int_0^\pi g_x\,\dd\theta
      =\beta\ell\sqrt f\,D(x).
\end{aligned}
  \label{eq:schmel-sph-polar-excess}
\end{equation}
Each pole contributes, at logarithmic order,
$(\beta\ell\sqrt f/2)\ln x$. Upon adding the bare sum
\eqref{eq:schmel-sph-bare-exact}, the action is exactly
\begin{equation}
  I^E_{\mathrm{g}}(r)
    =\beta\left[r(\sqrt f-1)+\frac{3m}{2}\right]
      +\beta\ell\sqrt f\,D(r/\ell).
  \label{eq:schmel-sph-pure-action-exact}
\end{equation}
At fixed $m$ and $\ell$, $\sqrt f=1-m/r+\Order(r^{-2})$ and
\begin{equation}
\begin{aligned}
  I^E_{\mathrm{g}}(r)
     ={}&\frac{\beta m}{2}
          +\beta\ell\left[\ln\frac{4r}{\ell}-1\right]\\
       &+\Order\!\left(\frac{\beta m^2}{r}
           +\frac{\beta\ell(m+\ell)}{r}\ln\frac r\ell\right).
\end{aligned}
  \label{eq:schmel-sph-pure-action-log}
\end{equation}
The factor $\sqrt f$ modifies subleading orders and does not remove
the logarithm. Thus, the renormalization defect consists of
the integrated contribution of the polar annuli, even though their
angular width tends to zero.

The functional~\eqref{eq:melvin-sph-pure-counterterm} coincides
with the one written in Eq.~(8)
of~\cite{Radu:2002}, where it is stated that it yields a
finite action in agreement with the background-subtraction
value.\footnote{The finiteness of the action is stated
in Ref.~\cite{Radu:2002} without specifying the boundary surface or
displaying the evaluation.}
Equations~\eqref{eq:melvin-sph-pure-log} and
\eqref{eq:schmel-sph-pure-action-log} show that this is not the case on
the complete spherical boundary, at fixed $\ell$ and without additional
subtractions. A finite value is obtained if only the leading term of
the fixed-angle expansion, $2r\sqrt f\,\sin\theta$, is integrated: it
contributes $\beta r\sqrt f$ and, added to
Eq.~\eqref{eq:schmel-sph-bare-exact}, gives $\beta m/2$ as $r\to\infty$.
The logarithm arises from the polar annuli, where that expansion is not
valid. The discrepancy is limited to the counterterm prescription and
does not affect the result $I^E=\beta m/2$ obtained
in~\cite{Radu:2002} through subtraction of the Melvin background.

The zero-field limit, $\ell\to\infty$ at fixed radius, lies
outside this expansion in $r/\ell\to\infty$. At zero field
the excess disappears and the seed counterterm is
recovered; that limit must not be inferred from the
asymptotic expressions at fixed $\ell$.

This difficulty motivates transporting the geometric functional of
the seed by means of the same transformation that generates the
magnetized geometry. That construction is developed in
Sec.~\ref{sec:harrison-construction}; it is not based on
a fit to the integrated value of the action.

\section{Construction of the counterterm via the Harrison transformation}
\label{sec:harrison-construction}

The Harrison transformation provides a relation between
the magnetized geometry and its seed that also extends to the
boundary terms. To make this transport precise, we first
present the symmetry of the reduced theory and the cancellation between
its bulk divergence and the Gibbons--Hawking--York term. We then
obtain the intrinsic identity that relates the seed counterterm
to that of the magnetized fields.

\subsection{Axial reduction and magnetizing transformation}
\label{subsec:harrison-generation}

We use the magnetizing transformation associated with the axial Killing vector
$\eta=\partial_\varphi$, of period $2\pi$
\cite{Harrison:1968,Ernst:1976,Ortaggio:2005}. We restrict ourselves
to the sector in which $\eta$ is hypersurface orthogonal, the fields
are independent of $\varphi$, and the potential is aligned with
$\dd\varphi$. Away from the axis,
\begin{equation}
\begin{aligned}
  \dd s^2&=\bar g_{ij}\dd x^i\dd x^j+V\dd\varphi^2,
  & A&=\psi\,\dd\varphi,\\
  V&=g(\eta,\eta),
  & F&=\partial_i\psi\,\dd x^i\wedge\dd\varphi,
  \qquad i,j=0,1,2,
\end{aligned}
  \label{eq:app-harrison-ansatz}
\end{equation}
The metric $\bar g$ has signature $(-++)$. The change of variables
\begin{equation}
  \gamma_{ij}=V\bar g_{ij},\qquad u=\ln(V/V_*),
  \label{eq:app-harrison-Einstein-frame}
\end{equation}
introduces the three-dimensional orbit metric $\gamma$; $V_*>0$
is constant and has the units of $V$. In this reduction,
$u$ denotes the logarithmic scalar, not the local coordinate used
to describe the poles in Sec.~\ref{sec:melvin-universe}.
We denote by $D$ and $\Box_\gamma$ the connection and the wave
operator of $\gamma$.

After integrating over the axial circle and separating a total divergence,
whose contribution we will retain in the next subsection, the
reduced action is
\begin{equation}
  I_{\mathrm{red}}=\frac{2\pi}{2\kappa}
   \int\dd^3x\,\sqrt{-\gamma}\left[
     R[\gamma]-\frac{\gamma^{ij}\partial_iV\partial_jV}{2V^2}
       -\frac{2\gamma^{ij}\partial_i\psi\partial_j\psi}{V}\right].
  \label{eq:harrison-reduced-action}
\end{equation}
The reduction steps and the resulting equations are kept
in Appendix~\ref{app:harrison-reduction}. Defining
$\zeta=\psi+\I\sqrt V$, the metric of the scalar sector takes the form
\begin{equation}
  \frac{(\dd V)^2}{2V^2}+\frac{2(\dd\psi)^2}{V}
       =2\frac{\dd\zeta\,\dd\bar\zeta}
                    {(\operatorname{Im}\zeta)^2}.
  \label{eq:harrison-target-metric}
\end{equation}
The combination~\eqref{eq:harrison-target-metric} is \emph{twice} the standard hyperbolic metric of the Poincaré upper half-plane, $d\zeta\,d\bar\zeta/(\operatorname{Im}\zeta)^2$, evaluated on the derivatives of the fields on the orbit space. The real linear fractional (Möbius) transformation
\begin{equation}
  \zeta\longmapsto\zeta'=\frac{\zeta}{1+b\zeta},\qquad
  \gamma_{ij}\longmapsto\gamma'_{ij}=\gamma_{ij},
  \qquad b\in\mathbb R,
  \label{eq:harrison-mobius}
\end{equation}
belongs to the one-parameter subgroup of $\mathrm{PSL}(2,\mathbb R)$ represented by the matrix $\bigl(\begin{smallmatrix}1&0\\ b&1\end{smallmatrix}\bigr)$, and is an isometry of that metric. Indeed, with
\begin{equation}
  \Lambda=|1+b\zeta|^2=(1+b\psi)^2+b^2V>0,
  \label{eq:app-harrison-complex-map}
\end{equation}
one has
\begin{equation}
  \dd\zeta'=\frac{\dd\zeta}{(1+b\zeta)^2},\qquad
  \operatorname{Im}\zeta'=\frac{\operatorname{Im}\zeta}{\Lambda},
  \qquad
  \frac{\dd\zeta'\,\dd\bar\zeta'}{(\operatorname{Im}\zeta')^2}
       =\frac{\dd\zeta\,\dd\bar\zeta}{(\operatorname{Im}\zeta)^2}.
  \label{eq:app-harrison-isometry}
\end{equation}
Since $\gamma$ remains fixed, the reduced density without total divergences
is pointwise invariant and its equations of motion are
preserved. In real variables, the map is
\begin{equation}
  \bar g'_{ij}=\Lambda^2\bar g_{ij},\qquad
  V'=\frac V{\Lambda^2},\qquad
  \psi'=\frac{\psi+b(\psi^2+V)}{\Lambda}.
  \label{eq:app-harrison-general-map}
\end{equation}
In particular, $\gamma'_{ij}=V'\bar g'_{ij}=V\bar g_{ij}=\gamma_{ij}$, although $\bar g'_{ij}\neq\bar g_{ij}$ for a nontrivial magnetization. The transformation preserves the equations of the reduced system. Whether the \emph{complete action} with its boundary terms is also invariant requires restoring exactly the terms that the reduction had separated, which we do next.

\subsection{Transformation of the action and boundary terms}

The symmetry of the reduced action is not sufficient to establish the
behavior of the four-dimensional action with boundary.
The exact reduction retains the total divergence
\begin{equation}
 \sqrt{-g}(R-F^2)=\sqrt{-\gamma}
  \left[R[\gamma]-\frac{(DV)^2}{2V^2}
             -\frac2V(D\psi)^2+\Box_\gamma u\right].
  \label{eq:app-harrison-exact-density}
\end{equation}
Under the Harrison transformation, $u'=u-2\ln\Lambda$, and therefore
\begin{equation}
  \sqrt{-g'}(R[g']-F'^2)-\sqrt{-g}(R[g]-F^2)
     =-2\partial_i\left(\sqrt{-\gamma}\,
                         D^i\ln\Lambda\right).
  \label{eq:app-harrison-bulk-divergence}
\end{equation}
This equality does not require imposing the field equations.

Consider a smooth, timelike, axially invariant boundary
that does not intersect the axis. Let $\mathcal B$ be the
three-dimensional orbit domain, $s$ the metric induced by $\gamma$
on its boundary, $\nu^i$ the outward unit spacelike normal,
and $k_\gamma$ its extrinsic curvature. With the convention
\eqref{eq:extrinsic-curvature-convention},
\begin{equation}
\begin{aligned}
  n^i&=\sqrt V\,\nu^i,\qquad n^\varphi=0,
  &\sqrt{-h}&=V^{-1/2}\sqrt{-s},\\
  K&=\sqrt V\left(k_\gamma-\tfrac12\nu^iD_i u\right),
  &\sqrt{-h}\,K&=\sqrt{-s}
                    \left(k_\gamma-\tfrac12\nu^iD_i u\right).
\end{aligned}
  \label{eq:app-harrison-GHY-reduction}
\end{equation}
After integrating over the axial circle, the divergence in
Eq.~\eqref{eq:app-harrison-exact-density} contributes the flux
\[
  \frac{2\pi}{2\kappa}\int_{\partial\mathcal B}\dd^2x\,
                                  \sqrt{-s}\,\nu^iD_i u.
\]
Its coefficient is opposite to that of the corresponding
Gibbons--Hawking--York contribution. The sum becomes
\begin{equation}
\begin{aligned}
  I_{\mathrm{bulk+GH}}
    ={}&\frac{2\pi}{2\kappa}\int_{\mathcal B}\dd^3x\,
      \sqrt{-\gamma}\left[R[\gamma]-\frac{(DV)^2}{2V^2}
                                   -\frac2V(D\psi)^2\right]\\
       &+\frac{2\pi}{\kappa}\int_{\partial\mathcal B}\dd^2x\,
                                        \sqrt{-s}\,k_\gamma.
\end{aligned}
  \label{eq:app-harrison-action-with-boundary}
\end{equation}
For the same orbit domain and the same axial period,
this expression is invariant: $\gamma$, $s$, and $k_\gamma$
remain fixed. The statement applies to the smooth timelike
boundaries specified; it does not include spacelike caps
or joint terms.

On the axis, $V=0$ and $\gamma$ degenerates even though the
four-dimensional geometry is regular. The fluxes of the reduction
cannot be discarded separately. The treatment by means of a regulating
tube, with the cancellations carried out before removing it,
is kept in Appendix~\ref{app:harrison-GHY}. For the
regular seeds in smooth axial gauge and their images in this
work, $\Lambda=1+\Order(\rho^2)$, and the comparison extends
to the axis as explained there.

\subsection{Intrinsic boundary invariant}
\label{subsec:harrison-counterterm}

The transformation of the counterterm requires an additional intrinsic
identity, independent of the cancellation of the boundary terms of the
action.
We pass to Euclidean signature: the metrics $q$ and $s$ of this
subsection are positive definite, unlike the Lorentzian induced
metric used in the previous subsection.

Consider first general axial Euclidean data, not
necessarily arising from a solution, and write, away from
the axis,
\begin{equation}
\begin{aligned}
  \dd s^2_{\Sigma,E}&=h^E_{ab}\dd x^a\dd x^b
                      =q_{AB}\dd x^A\dd x^B+V\dd\varphi^2,\\
  A|_\Sigma&=\psi\,\dd\varphi,\qquad
  s_{AB}=Vq_{AB},\qquad A,B=1,2.
\end{aligned}
  \label{eq:harrison-boundary-reduced-data}
\end{equation}
We assume $q$ positive definite, $V=h_E(\eta,\eta)>0$, and fields independent of
$\varphi$. If these data are induced from the bulk, $s$ is
precisely the metric induced by $\gamma$ on the boundary of the
orbit space. The indices $A,B$ refer to this
two-dimensional base, which is distinct from the spatial sections with metric
$\sigma_{ij}$ used for the Brown--York charges. We denote
by $D_q,D_s$ the connections of the auxiliary metrics $q,s$,
by $D_q^2,D_s^2$ their Laplacians, and by $R[q],R[s]$ their
scalar curvatures; we reserve $\cR$ for the three-dimensional
boundary. In particular,
$|D_s u|_s^2=s^{AB}\partial_Au\partial_Bu$, where
$u=\ln(V/V_*)$ and $V_*>0$ is a constant with the units of $V$.
The tangential contraction
of the metric of the reduced fields is
\begin{equation}
  \mathcal T_s(V,\psi)
    =\frac{s^{AB}\partial_AV\partial_BV}{2V^2}
       +\frac{2s^{AB}\partial_A\psi\partial_B\psi}{V}
    =\tfrac12|D_su|_s^2
       +\frac2V|D_s\psi|_s^2.
  \label{eq:harrison-boundary-target-contraction}
\end{equation}
Here $\mathcal T_s$ is a scalar quantity, not an energy
tensor. Its coefficients are the same as in
Eq.~\eqref{eq:harrison-reduced-action}, fixed by the
$R-F^2$ normalization of the theory.

For a one-dimensional fiber and a two-dimensional base one has
the geometric identities
\begin{equation}
  \cR[q+v^2\dd\varphi^2]=R[q]-2v^{-1}D_q^2 v,
  \label{eq:harrison-warped-curvature}
\end{equation}
\begin{equation}
  R[e^{2\omega}q]
       =e^{-2\omega}(R[q]-2D_q^2\omega),
  \qquad D_{e^{2\omega}q}^2v=e^{-2\omega}D_q^2 v.
  \label{eq:harrison-base-conformal}
\end{equation}
Here $v$ is a local auxiliary quantity. Applying these identities to
$q=V^{-1}s$ and $v=\sqrt V$ gives
\begin{equation}
\begin{aligned}
  R[q]&=V\bigl(R[s]+D_s^2 u\bigr),\\
  2V^{-1/2}D_q^2\sqrt V
    &=V\bigl(D_s^2 u+\tfrac12|D_su|_s^2\bigr),\\
  \mathcal F^2&=\frac2Vq^{AB}\partial_A\psi\partial_B\psi
                    =2|D_s\psi|_s^2.
\end{aligned}
  \label{eq:harrison-intrinsic-reduction}
\end{equation}
We define $\mathcal F=\dd(A|_\Sigma)$ and raise its indices
with $h_E$. This invariant contains only the induced field:
it is not the restriction of the four-dimensional scalar $F^2$, which
also includes components normal to $\Sigma$. The two
contributions $D_s^2 u$ cancel pointwise, so that
\begin{equation}
\quad
  X\equiv\cR[h_E]-\mathcal F^2=V\mathcal Y,
  \qquad \mathcal Y\equiv R[s]-\mathcal T_s(V,\psi).
  \quad
  \label{eq:harrison-intrinsic-reduced-invariant}
\end{equation}
This cancellation relies on the base being two-dimensional and does not discard
any divergence. Both $R[s]$ and the gradients present in
$\mathcal T_s$ are intrinsic: $\mathcal Y$ is not obtained
simply by restricting the reduced bulk scalar to
$\Sigma$. The identity holds pointwise, before any integration
by parts.

Under the general map~\eqref{eq:app-harrison-general-map},
$q'=\Lambda^2q$ and $V'=V/\Lambda^2$, so that $s'=s$.
The isometry~\eqref{eq:app-harrison-isometry} likewise implies
$\mathcal T_s(V',\psi')=\mathcal T_s(V,\psi)$. Consequently,
\begin{equation}
  \mathcal Y'=\mathcal Y,\qquad X'=\Lambda^{-2}X,
  \qquad \sqrt{h'_E}=\Lambda\sqrt{h_E},\qquad \Lambda>0.
  \label{eq:harrison-intrinsic-covariance}
\end{equation}
The connection with the square root of the counterterm is made explicit in
\begin{equation}
  \sqrt{h_E}\sqrt{2X}
      =\sqrt{s}\sqrt{2\mathcal Y}
      =\sqrt{h'_E}\sqrt{2X'},\qquad X\geq0,
  \label{eq:harrison-density-reduced-invariant}
\end{equation}
where the nonnegative root is taken. Upon integrating over the axial circle,
the first expression is integrated with $\dd^2x\,\dd\varphi$ and
the reduced expression acquires the factor $2\pi$. The identity
of Appendix~\ref{app:harrison-boundary-identity} is thus obtained
directly from the scalar structure of the reduction.

\subsection{Transport from neutral seeds and spherical application}

\paragraph{Generation of the geometries.}
We distinguish the fields of the vacuum seed with a hat:
\begin{equation}
  \dd\hat s^2=\hat g^{\perp}_{ij}\dd x^i\dd x^j+G\dd\varphi^2,
  \qquad G=\hat g(\eta,\eta),\qquad \hat A=0.
  \label{eq:harrison-vacuum-seed}
\end{equation}
The indices $i,j$ run over the three coordinates of the orbit
space; $G$ is the squared axial norm of the seed.
In the variables of the axial reduction,
$\gamma_{ij}=G\hat g^{\perp}_{ij}$ and $\zeta=\I\sqrt G$.
Substituting into Eq.~\eqref{eq:harrison-mobius}, with $b=1/\ell$ and
$\ell>0$, we obtain
\begin{equation}
  \zeta'=\frac{G/\ell+\I\sqrt G}{1+G/\ell^2},
  \qquad \Lambda=1+\frac G{\ell^2},
  \label{eq:harrison-neutral-mobius}
\end{equation}
and, therefore,
\begin{equation}
\begin{aligned}
  \dd s^2&=\Lambda^2\hat g^{\perp}_{ij}\dd x^i\dd x^j
                          +\Lambda^{-2}G\dd\varphi^2,\\
  A&=\frac{G}{\ell\Lambda}\dd\varphi,
  \qquad F=\frac{\dd G\wedge\dd\varphi}{\ell\Lambda^2}.
\end{aligned}
  \label{eq:harrison-neutral-map}
\end{equation}
The relation between the potential and the axial norm is part of the
map: no test field is added to a fixed geometry, as
in the Wald approximation~\cite{Wald:1974}. The
factors $\Lambda^2$ and $\Lambda^{-2}$ incorporate its gravitational effect.

For the Minkowski seed in cylindrical coordinates,
\begin{equation}
  \dd\hat s^2=-\dd t^2+\dd\rho^2+\dd z^2
                       +\rho^2\dd\varphi^2,
  \qquad G=\rho^2,
  \label{eq:harrison-Minkowski-seed}
\end{equation}
the result is
\begin{equation}
  \dd s^2=\Lambda^2(-\dd t^2+\dd\rho^2+\dd z^2)
                   +\frac{\rho^2}{\Lambda^2}\dd\varphi^2,
  \qquad \Lambda=1+\frac{\rho^2}{\ell^2}.
  \label{eq:harrison-Melvin-cylindrical}
\end{equation}
In the meridional and azimuthal orthonormal frame,
$F_{\hat\rho\hat\varphi}=2/(\ell\Lambda^2)$, so that the
field on the axis is $B_0=2/\ell=2b$. Finally,
$\rho=r\sin\theta$, $z=r\cos\theta$ allow the
result to be written in the spherical form~\eqref{eq:harrison-Melvin-spherical}.
This is the Melvin solution that we will use. If, instead,
one chooses the seed
\begin{equation}
  \dd\hat s^2=-f\dd t^2+f^{-1}\dd r^2+r^2\dd\theta^2
                          +r^2\sin^2\theta\dd\varphi^2,
  \qquad f=1-\frac{2m}{r},
  \label{eq:harrison-Schwarzschild-seed}
\end{equation}
the same $G=r^2\sin^2\theta$ produces Schwarzschild--Melvin:
the potential and $\Lambda$ keep their form and $f$ remains in
the block orthogonal to the axis. Both solutions thus arise from a
single map. On a regular axis, $G=\rho^2+\Order(\rho^4)$,
$\Lambda=1+\Order(\rho^2)$, and $A_\varphi=\Order(\rho^2)$;
the map preserves axial regularity and the period $2\pi$.
In particular, if $x=\rho\cos\varphi$ and $y=\rho\sin\varphi$,
$\rho^2\dd\varphi=x\dd y-y\dd x$ is a smooth form in
transverse Cartesian coordinates.

\paragraph{Induced data of the seed.}
We now specialize the initial data to $\hat A=0$, and
reserve the hat for the seed and the unprimed symbols
for the magnetized configuration. The boundary metric
of the seed is
\begin{equation}
  \dd\hat s^2_{\Sigma,E}
       =\hat q_{AB}\dd x^A\dd x^B+G\dd\varphi^2,
  \qquad A,B=1,2,
  \label{eq:harrison-seed-boundary}
\end{equation}
with $\hat q$ positive definite and $G>0$. In this proof, $\hat q$ and
$G$ are arbitrary functions independent of $\varphi$; they are not
required to be restrictions of a vacuum solution. The map
induced by Eq.~\eqref{eq:harrison-neutral-map} is
\begin{equation}
\begin{aligned}
  \dd s^2_{\Sigma,E}
    &=\Lambda^2\hat q_{AB}\dd x^A\dd x^B
                          +\Lambda^{-2}G\dd\varphi^2,\\
  A\big|_\Sigma&=\frac{G}{\ell\Lambda}\dd\varphi,
  \qquad \Lambda=1+\frac G{\ell^2},
  \qquad \sqrt{h_E}=\Lambda\sqrt{\hat h_E}.
\end{aligned}
  \label{eq:harrison-boundary-map}
\end{equation}
Since the induced field of the seed vanishes,
$\hat X=\hat{\cR}\equiv\cR[\hat h_E]$. The covariance
\eqref{eq:harrison-intrinsic-covariance} reduces to
\begin{equation}
  \boxed{\ \cR[h_E]-\mathcal F_{ab}\mathcal F^{ab}
                    =\Lambda^{-2}\hat{\cR}\ }.
  \label{eq:harrison-curvature-identity}
\end{equation}
The explicit check in the seed variables,
Eqs.~\eqref{eq:harrison-curvature-transformation} and
\eqref{eq:harrison-curvature-Maxwell-cancellation}, is kept
in Appendix~\ref{app:harrison-boundary-identity}.
On a regular axis,
$G=\rho^2+\Order(\rho^4)$ and the magnetized potential is
$\Order(\rho^2)\dd\varphi$ in a smooth gauge. The three-dimensional
identities then extend by continuity, even though the
reduced metric $s$ degenerates there. No integration by parts has been
performed at the poles, nor has any flux of the axial reduction been assumed to vanish.

\paragraph{Transport of the density and selection of the coefficient.}
If $\hat{\cR}\geq0$, the nonnegative root and
$\sqrt{h_E}=\Lambda\sqrt{\hat h_E}$, Eq.~\eqref{eq:harrison-boundary-map}, give
\begin{equation}
  \boxed{\ \sqrt{h_E}\sqrt{2(\cR[h_E]-\mathcal F^2)}
      =\sqrt{\hat h_E}\sqrt{2\cR[\hat h_E]}\ }.
  \label{eq:harrison-counterterm-density}
\end{equation}
The right-hand side is the density of the three-dimensional intrinsic
counterterm used for asymptotically flat regulators of
type $S^2\times\mathbb R$, or $S^2\times S^1$ after identifying
the Euclidean time
\cite{Mann:1999,KrausLarsenSiebelink,AstefaneseiMannStelea:2007}.
The reference by Lau uses the curvature of the two-dimensional
spatial section and constitutes a related antecedent
\cite{Lau:1999}; here the identity refers explicitly
to the three-dimensional curvature of $\hat h_E$. The choice of that
functional presupposes a suitable class of boundaries in the seed;
its use does not constitute a universal prescription for every
asymptotically flat geometry~\cite{KrausLarsenSiebelink}. For those
geometries there also exists the Mann--Marolf counterterm, with
its associated boundary tensor~\cite{MannMarolf:2006,MannMarolfMcNeesVirmani:2008}.

Consequently, the geometric functional of the seed admits the
intrinsic extension to the magnetized fields
\begin{equation}
\begin{aligned}
  I^E_{\mathrm{ct}}[h_E,A]
      &\equiv\frac1\kappa\int_\Sigma\dd^3x\,
                \sqrt{h_E}\sqrt{2(\cR[h_E]-\mathcal F^2)},\\
  I^E_{\mathrm{ct}}[\mathcal H_\ell(\hat h_E)]
      &=I^E_{\mathrm{ct,g}}[\hat h_E],
  \qquad
  I^E_{\mathrm{ct,g}}[\hat h_E]
       =\frac1\kappa\int_\Sigma\dd^3x\,
                    \sqrt{\hat h_E}\sqrt{2\hat{\cR}},
\end{aligned}
  \label{eq:harrison-counterterm-transfer}
\end{equation}
where $\mathcal H_\ell$ denotes the data map
$\hat h_E\mapsto(h_E,A|_\Sigma)$ of Eq.~\eqref{eq:harrison-boundary-map}.
This is the precise sense in which the new counterterm derives
from $\sqrt{2\hat{\cR}}$ through the Harrison transformation: its composition with the data
map reproduces the seed functional. The symmetry preserves
$\sqrt{h_E}\sqrt{2X}$ and not the scalar curvature separately.
Evaluating
$\sqrt{h_E}\sqrt{2\cR[h_E]}$ directly retains the additional electromagnetic
term of Eq.~\eqref{eq:harrison-curvature-identity} and yields a different
functional. This distinction is what motivates the present construction.

\paragraph{Application to the spherical boundaries.}
For Minkowski or Schwarzschild, the Euclidean induced metric at
$r=\mathrm{const.}$ is
\begin{equation}
\begin{aligned}
  \dd\hat s^2_{\Sigma,E}
     &=f(r)\dd\tau^2+r^2\dd\theta^2+r^2\sin^2\theta\dd\varphi^2,
  &\hat{\cR}&=\frac2{r^2},\\
  \sqrt{\hat h_E}&=r^2\sqrt f\,\sin\theta,
  &\cR[h_E]-\mathcal F^2&=\frac2{r^2\Lambda^2},
\end{aligned}
  \label{eq:harrison-spherical-boundary}
\end{equation}
with $f=1$ or $f=1-2m/r>0$. The curvature of the seed does not depend
on $f$, which is constant on each cutoff. Therefore,
\begin{equation}
  I^E_{\mathrm{ct}}(r)
    =\frac{\beta(2\pi)}\kappa
         \int_0^\pi\dd\theta\,2r\sqrt f\,\sin\theta
    =\beta r\sqrt f.
  \label{eq:harrison-spherical-counterterm}
\end{equation}
The equality is exact at finite cutoff and includes the poles by
continuity. The magnetized density is exactly
$2r\sqrt f\,\sin\theta$, so that the complete sphere is integrated
without an expansion of $\Lambda$ at fixed angle. This explains why
the transport does not lose the polar regions responsible for the
failure of the purely geometric counterterm.

\paragraph{Invariance of the renormalized action.}
The transport identity~\eqref{eq:harrison-counterterm-density} can be
combined with the invariance of the sum of the bulk and
Gibbons--Hawking--York terms,
Eq.~\eqref{eq:app-harrison-action-with-boundary}, extended to the axis
as in Appendix~\ref{app:harrison-GHY}. For the regulated regions
considered here, with the same orbit domain, Euclidean period, and
axial period, the result is
\begin{equation}
\left.\left(I^E_{\mathrm{bulk}}+I^E_{\mathrm{GH}}
+I^E_{\mathrm{ct}}\right)\right|_{\mathrm{magnetized}}
=\left.\left(I^E_{\mathrm{bulk}}+I^E_{\mathrm{GH}}
+I^E_{\mathrm{ct,g}}\right)\right|_{\mathrm{seed}}
\label{eq:harrison-action-invariance}
\end{equation}
at every finite cutoff: the action of the magnetized solution,
renormalized with $\sqrt{2(\cR-\mathcal F^2)}$, equals that of its seed,
renormalized with $\sqrt{2\hat{\cR}}$. For the Minkowski and
Schwarzschild seeds the right-hand side is $0$ and
$\beta[r(\sqrt f-1)+3m/2]$, respectively. The direct evaluations of
Sec.~\ref{sec:euclidean-action},
Eqs.~\eqref{eq:melvin-sph-renormalized-exact} and
\eqref{eq:schmel-sph-renormalized-exact}, verify this equality
explicitly. It accounts for the independence of the action of the
external field noted in~\cite{Radu:2002,Ortaggio:2005}.

The invariance concerns the value of the action. It does not extend to
its functional derivatives with respect to independent boundary data,
so that the boundary tensor, the current, and the charges are not fixed
by it; they are obtained in Secs.~\ref{sec:variational-principle}
and~\ref{sec:brown-york-energy}.

\paragraph{Inherited domain of validity.}
The construction does not remove the conditions under which the seed
counterterm applies; it transports them. The functional
$\sqrt{2\hat{\cR}}$ renormalizes asymptotically flat seeds on a suitable
class of boundaries, such as the $S^1\times S^2$ regulator used here,
and is not a universal prescription for every asymptotically flat
geometry~\cite{KrausLarsenSiebelink,MannMarolf:2006}.
Equation~\eqref{eq:harrison-action-invariance} shows that, whenever a
seed in the hypersurface-orthogonal axial sector has a finite
renormalized action with $\sqrt{2\hat{\cR}}$ on a given regulator, its
Harrison image has a finite renormalized action with
$\sqrt{2(\cR-\mathcal F^2)}$ on the regulator with the same orbit
domain, and the two values coincide at each cutoff. Dirichlet
stationarity at finite cutoff holds on both sides, since both
counterterms are intrinsic. In this sense, the domain of validity of
the new counterterm is the Harrison image of that of the seed
counterterm. The map does not transport the asymptotic variational
problem for general perturbations: it acts only on axial data and does
not map the space of independent variations onto itself, as discussed
next.

The coefficient of $\mathcal F^2$ is also singled out by the complete
integral. For the family of densities
$\sqrt{h_E}\sqrt{2(\cR-a\mathcal F^2)}$ with fixed $a<1$,
Appendix~\ref{app:prd-esfera-pura} shows that the action on Melvin
retains the excess $\beta\ell\sqrt{1-a}\,\ln(r/\ell)$ at leading order,
Eq.~\eqref{eq:polar-family-D-asymptotic}, whereas for $a=1$ the excess
vanishes identically.

\paragraph{Scope of the transport.}
The equality of densities relates different boundary data:
the Harrison transformation acts simultaneously on the metric and the potential.
It does not identify Dirichlet problems with the same data, nor does it
allow the independent variations of the Brown--York
functional to be restricted to the Harrison orbit. The intrinsic extension
of Eq.~\eqref{eq:harrison-counterterm-transfer} must be varied with
those independent sources, in the domain
$X=\cR[h_E]-\mathcal F^2>0$ specified in
Eq.~\eqref{eq:melvin-sph-E-variation-domain}. At $X=0$ the
equality of values may persist, but the general derivative of
the square root is not guaranteed.

The counterterm is thus obtained by transport of the geometric
functional of the seed, without fitting its coefficient to an
integrated value of the action or of the energy. The identity~\eqref{eq:harrison-intrinsic-covariance} is not restricted
to neutral seeds: it holds pointwise for any axially symmetric boundary
data with hypersurface-orthogonal $\eta$ and potential aligned with
$\dd\varphi$, including seeds with $\psi\neq0$, for which the seed
functional is itself $\sqrt{2(\hat{\cR}-\hat{\mathcal F}^2)}$.
Appendix~\ref{app:harrison-charged-seeds} checks it explicitly for a
magnetized seed and for a magnetically charged seed. Its global use for
such seeds requires, in addition, the analysis of the gauge and of the
axial period described in Appendix~\ref{app:harrison-neutral-axis}: for
a magnetically charged seed the potential cannot vanish at both poles. Rotating seeds lie outside
this sector: for Kerr--Melvin, $g_{i\varphi}\neq0$ and the potential
acquires components $A_i$, so that the reduction must be enlarged, and
the invariance of the density has not been established there. The
identity establishes neither the uniqueness of the counterterm nor a
universal prescription for asymptotically Melvin geometries.

\section{Variational principle and boundary responses}
\label{sec:variational-principle}

The Harrison construction fixes the counterterm, but not the boundary
responses of the complete action.
Since the counterterm is intrinsic, it introduces no normal derivatives
of the variations, and the Dirichlet problem remains stationary at
finite cutoff, as for the Gibbons--Hawking--York action. The content
of this section is therefore the boundary responses: the Brown--York
tensor, the current conjugate to the tangential potential, and the Ward
identities that they satisfy. These responses are not determined by the
value of the action on the solutions.

\subsection{Independent data and domain of the functional}

We consider the outer boundaries of fixed radius of
Sec.~\ref{sec:melvin-universe}, with the convention for the normal
and the extrinsic curvature of Eq.~\eqref{eq:extrinsic-curvature-convention}.
In Lorentzian signature, the spatial section of the boundary
is closed. The variations are smooth, regular on the axis,
and have compact support in $t$, so that the temporal
endpoints do not contribute to the first variation. In the application
to Melvin, regularity at the origin is also required. The poles
do not constitute additional edges of the spatial section.

The complete action that we vary is
\begin{equation}
  I_{\mathrm{ren}}
  =\frac1{2\kappa}\int_{\cM}\dd^4x\,\sqrt{-g}\,(R-F^2)
    +\frac1\kappa\int_\Sigma\dd^3x\,\sqrt{-h}\,K
    -\frac1\kappa\int_\Sigma\dd^3x\,\sqrt{-h}\,U,
  \label{eq:melvin-sph-BY-L-action}
\end{equation}
with the intrinsic definitions
\begin{equation}
  U=\sqrt{2(\cR-\mathcal F^2)},\qquad
  \Xi=U^{-1},\qquad
  \mathcal F_{ab}=h_a{}^\mu h_b{}^\nu F_{\mu\nu},\qquad
  \mathcal F^2=h^{ac}h^{bd}\mathcal F_{ab}\mathcal F_{cd}.
  \label{eq:melvin-sph-BY-intrinsic-definitions}
\end{equation}
The induced field is $\mathcal F=\dd(A|_\Sigma)$; its indices
are raised with $h_{ab}$, not with a metric fixed beforehand.
We denote by $D_a$ the connection compatible with $h_{ab}$ and
by $D^2=D_aD^a$ its second-order operator. The domain of
the positive branch is expressed by
\begin{equation}
  X=\cR-\mathcal F^2,\qquad U=\sqrt{2X},\qquad
  \qquad X>0.
  \label{eq:melvin-sph-E-variation-domain}
\end{equation}
The same notation will be used in Euclidean signature, with
contractions performed with $h_E$. The equality of values
of the counterterm can be extended to $X=0$, but the general
derivative of the square root is not guaranteed there.

The induced metric and the tangential potential are varied
independently, keeping the cutoff fixed. The variations are not
restricted to the Harrison orbit or to the exact families
parametrized by $m$ and $\ell$. In particular, merely differentiating
$I^E_{\mathrm{ren}}(r)$ along the family parametrized
by $m$ does not determine the boundary tensor. The identities
$U=2/(r\Lambda)$ and $\Xi=r\Lambda/2$ are used after
varying, not as substitutions in the functional.

The counterterm contains second tangential derivatives of the
metric through $\cR$ and first tangential derivatives
of the potential through $\mathcal F$. It contains neither $K_{ab}$
nor $F_{na}$, so its variation does not introduce new
normal derivatives of the variations. The tangential derivatives
are transferred to their coefficients by intrinsic integrations
by parts. The Gibbons--Hawking--York term cancels the normal derivatives
of the metric variation of the Einstein--Hilbert term, whereas the Maxwell term
leaves a contribution proportional to $\delta A_a$
\cite{GibbonsHawking,JubbEtAl:2017}.

\subsection{Complete first variation}

We impose the Einstein--Maxwell equations of motion
on the background, not on its variations. The joint contribution of the
bulk and Gibbons--Hawking--York terms is
\begin{equation}
\begin{aligned}
  \left.\delta I_{\mathrm{bulk+GH}}\right|_{\mathrm{EOM}}
  ={}&\frac1{2\kappa}\int_\Sigma\dd^3x\,\sqrt{-h}\,
          (K_{ab}-Kh_{ab})\delta h^{ab}\\
    &-\frac2\kappa\int_\Sigma\dd^3x\,\sqrt{-h}\,
          n_\mu F^{\mu a}\delta A_a.
\end{aligned}
  \label{eq:melvin-sph-BY-bare-variation}
\end{equation}
The index $a$ of the second term is tangential. The factor
$2/\kappa$ corresponds to the normalization $-F^2/(2\kappa)$
of the Maxwell action.

To vary the counterterm we use
\begin{equation}
\begin{aligned}
  \delta\sqrt{-h}&=-\frac12\sqrt{-h}\,h_{ab}\delta h^{ab},\\
  \delta U&=\Xi(\delta\cR-\delta\mathcal F^2),\\
  \delta\cR&=\mathcal R_{ab}\delta h^{ab}
          +(h_{ab}D^2-D_aD_b)\delta h^{ab},\\
  \delta\mathcal F^2
       &=2\mathcal F_{ac}\mathcal F_b{}^c\delta h^{ab}
            +4\mathcal F^{ab}D_a\delta A_b.
\end{aligned}
  \label{eq:melvin-sph-BY-variation-identities}
\end{equation}
The integrations by parts are intrinsic to $\Sigma$.
They generate no terms at the poles, and the contributions of the
temporal endpoints vanish because of the support of the variations.
Defining
\begin{equation}
  \mathcal C_{ab}
  =-\frac U2h_{ab}
       +\Xi(\mathcal R_{ab}-2\mathcal F_{ac}\mathcal F_b{}^c)
       +(h_{ab}D^2-D_aD_b)\Xi,
  \label{eq:melvin-sph-BY-ct-coefficient}
\end{equation}
one obtains
\begin{equation}
\begin{aligned}
  \delta I_{\mathrm{ct}}
   ={}&-\frac1\kappa\int_\Sigma\dd^3x\,\sqrt{-h}\,
                 \mathcal C_{ab}\delta h^{ab}\\
      &-\frac4\kappa\int_\Sigma\dd^3x\,\sqrt{-h}\,
                 D_a(\Xi\mathcal F^{ab})\delta A_b.
\end{aligned}
  \label{eq:melvin-sph-BY-ct-variation}
\end{equation}
The electromagnetic dependence of the counterterm modifies both
the metric response and the response conjugate to $A_a$.
Adding all the contributions,
\begin{equation}
  \left.\delta I_{\mathrm{ren}}\right|_{\mathrm{EOM}}
   =-\frac12\int_\Sigma\dd^3x\,\sqrt{-h}\,
                        \tau^{\mathrm{ren}}_{ab}\delta h^{ab}
     +\int_\Sigma\dd^3x\,\sqrt{-h}\,
                        \mathcal J^a_{\mathrm{ren}}\delta A_a,
  \label{eq:melvin-sph-BY-total-variation}
\end{equation}
where
\begin{equation}
\begin{aligned}
  \tau^{\mathrm{ren}}_{ab}
   &=-\frac2{\sqrt{-h}}\frac{\delta I_{\mathrm{ren}}}{\delta h^{ab}}
    =-\frac1\kappa(K_{ab}-Kh_{ab})+\frac2\kappa\mathcal C_{ab},\\
  \mathcal J^a_{\mathrm{ren}}
   &=\frac1{\sqrt{-h}}\frac{\delta I_{\mathrm{ren}}}{\delta A_a}
    =-\frac2\kappa n_\mu F^{\mu a}
       -\frac4\kappa D_b(\Xi\mathcal F^{ba}).
\end{aligned}
  \label{eq:melvin-sph-BY-responses}
\end{equation}
The first definition follows the
Brown--York procedure~\cite{BrownYork}. The second defines the
renormalized electromagnetic response, which is distinct from the spatial
momentum density $j_i$ that enters the quasilocal charges.
Both responses are identified before imposing Dirichlet
conditions.

An equivalent form of the tensor is obtained by introducing
$\mathcal T_{ab}=2\mathcal F_{ac}\mathcal F_b{}^c-h_{ab}\mathcal F^2$.
This algebraic boundary combination is not the Maxwell
energy--momentum tensor in the bulk. Since
$U/2=\Xi(\cR-\mathcal F^2)$,
\begin{equation}
\begin{aligned}
  \tau^{\mathrm{ren}}_{ab}
   ={}&-\frac1\kappa(K_{ab}-Kh_{ab})\\
      &+\frac2\kappa\left[
        \Xi(\mathcal R_{ab}-h_{ab}\cR-\mathcal T_{ab})
          +(h_{ab}D^2-D_aD_b)\Xi\right].
\end{aligned}
  \label{eq:brown-york-spherical}
\end{equation}
The derivatives of $\Xi$ are part of the metric response.
Substituting beforehand the expression of $U$ on a solution
would eliminate variations of $\cR$ and $\mathcal F^2$, including
the Hessian and current terms.

\subsection{Dirichlet conditions, Ward identities, and scope}

As anticipated, the first
variation~\eqref{eq:melvin-sph-BY-total-variation} shows that fixing
the induced data suffices for stationarity at finite cutoff:
\begin{equation}
  \delta h^{ab}\big|_\Sigma=0,\qquad
  \delta A_a\big|_\Sigma=0
  \quad\Longrightarrow\quad
  \left.\delta I_{\mathrm{ren}}\right|_{\mathrm{EOM,\,Dirichlet}}=0.
  \label{eq:melvin-sph-BY-Dirichlet}
\end{equation}
The tangential potential is fixed in a regular gauge, and the
admissible gauge transformations preserve those data.
Keeping $\ell$ fixed identifies the external field within
the exact family, but it does not replace the functional conditions
on $(h_{ab},A_a)$ for general perturbations.
It is not necessary to fix in addition $K_{ab}$ or the conjugate momenta,
nor to require the responses to vanish.

The intrinsic contribution to the electromagnetic response does not
constitute a Legendre transformation of the Maxwell term, nor does it change
the ensemble to one of fixed normal flux
\cite{BradenBrownWhitingYork}. In Euclidean signature
the data~\eqref{eq:melvin-sph-E-Dirichlet-data} are fixed, with $r$,
$\beta$, and $\ell$ fixed. The corresponding first variation
is presented in Eq.~\eqref{eq:melvin-sph-E-total-variation} and leads
to Eq.~\eqref{eq:melvin-sph-finite-Dirichlet}. There the boundary
$\Sigma\simeq S^1_\beta\times S^2$ is compact and has no edge;
the regularity of the potential at the poles is expressed by
Eq.~\eqref{eq:melvin-sph-polar-potential}.

The boundary identities associated with gauge invariance
and with the tangential diffeomorphisms of the complete action are,
on shell,
\begin{equation}
  D_a\mathcal J^a_{\mathrm{ren}}=0,\qquad
  D_a\tau^{\mathrm{ren}\,a}{}_{b}
                  =\mathcal F_{ba}\mathcal J^a_{\mathrm{ren}},
  \label{eq:melvin-sph-BY-Ward}
\end{equation}
The metric tensor can exchange momentum with the electromagnetic
source, so its divergence need not vanish
separately. The explicit polar check on
Schwarzschild--Melvin is kept in
Eq.~\eqref{eq:schmel-sph-BY-Ward-polar}. The additional conditions
for the conservation of the Killing-time charge are specified when
the charges are evaluated.

These results hold at finite cutoff and for the prescribed data. An
asymptotic variational principle for a general class of asymptotically
Melvin geometries would require specifying the admissible
perturbations, their falloff conditions, and the control of the square
root, since $X=2/(r^2\Lambda^2)$ tends to zero as the cutoff is removed
on the solutions considered. The vanishing of the responses on Melvin,
checked in Sec.~\ref{sec:brown-york-energy}, is a property of that
background and not a consequence of Dirichlet stationarity.

\section{Evaluation of the Euclidean action}
\label{sec:euclidean-action}

We evaluate the action on Melvin and Schwarzschild--Melvin with
the counterterm fixed by the Harrison transformation in
Sec.~\ref{sec:harrison-construction}. We use the same
functional in both geometries, without fitting its coefficients
to the value of the action and without background subtraction.
We distinguish the identities at finite cutoff from the limit that
removes the regulator.

\subsection{Prescription and evaluation on the solutions}

We keep the Euclidean conventions and the outward normal
of Sec.~\ref{sec:melvin-universe}. We denote by $\cM_r$
the regulated Euclidean region corresponding to each solution
and by $\Sigma$ its outer boundary. Before adding the
counterterm, the action is
\begin{equation}
  I^E_{\mathrm{bulk+GH}}(r)
  =-\frac{1}{2\kappa}\int_{\cM_r}\dd^4x\,\sqrt{g_E}\,(R-F^2)
   -\frac{1}{\kappa}\int_\Sigma\dd^3x\,\sqrt{h_E}\,K,
  \label{eq:melvin-sph-E-action}
\end{equation}
with $K_{ab}=h_a{}^\mu h_b{}^\nu\nabla_\mu n_\nu$, according to
Eq.~\eqref{eq:extrinsic-curvature-convention}~\cite{GibbonsHawking}.
To this action we add
\begin{equation}
  I^E_{\mathrm{ct}}[h_E,A]
    =\frac1\kappa\int_\Sigma\dd^3x\,\sqrt{h_E}\,
                  \sqrt{2(\cR-\mathcal F^2)},
  \label{eq:melvin-sph-counterterm}
\end{equation}
with the positive root. This is the intrinsic extension
obtained in Eq.~\eqref{eq:harrison-counterterm-transfer}, local and
invariant under $A\mapsto A+\dd\lambda$. The combination
$\cR-\mathcal F^2$ is positive on the cutoffs considered,
as shown by the invariants of
Sec.~\ref{sec:melvin-universe}. The variational problem,
with the induced metric and the tangential potential prescribed,
is that of Sec.~\ref{sec:variational-principle}.

In both solutions, the trace of the
Einstein--Maxwell equations gives $R=0$. The bulk invariant $F^2$
remains nonzero and must not be identified with
$\mathcal F^2$. The integrals are evaluated over the complete
sphere before performing radial expansions, keeping
$\ell>0$ fixed. The limitation of replacing the mixed functional
here by $\sqrt{2\cR}$ was established in
Sec.~\ref{sec:geometric-counterterm-limit}.

\subsection{Melvin: exact cancellation at finite cutoff}
\label{subsec:melvin-esfericas-renormalization}

The region is $0\leq\bar r\leq r$, with $r>0$ and arbitrary
Euclidean period $\beta$. The complete boundary is given by
Eq.~\eqref{eq:melvin-sph-boundary}; the origin and the poles
contribute no additional boundary components.
The bulk invariants~\eqref{eq:melvin-sph-bulk-invariants}
and the angular integrals~\eqref{eq:melvin-sph-angular-integrals}
yield the exact contributions
\eqref{eq:melvin-sph-bulk-exact} and
\eqref{eq:melvin-sph-GH-exact}. Their nonpolynomial terms
cancel before any limit is taken:
\begin{equation}
  I^E_{\mathrm{bulk}}(r)+I^E_{\mathrm{GH}}(r)=-\beta r.
  \label{eq:melvin-sph-bare-exact}
\end{equation}

For their part, the intrinsic invariants
\eqref{eq:melvin-sph-intrinsic-invariants} give
\begin{equation}
  \sqrt{2(\cR-\mathcal F^2)}=\frac{2}{r\Lambda},
  \qquad
  I^E_{\mathrm{ct}}(r)
   =\frac{\beta(2\pi)}\kappa
        \int_0^\pi\dd\theta\,2r\sin\theta=\beta r.
  \label{eq:melvin-sph-counterterm-exact}
\end{equation}
The sum of the three contributions is therefore
\begin{equation}
  I^E_{\mathrm{ren}}(r)
   =I^E_{\mathrm{bulk}}(r)+I^E_{\mathrm{GH}}(r)
                           +I^E_{\mathrm{ct}}(r)=0
  \label{eq:melvin-sph-renormalized-exact}
\end{equation}
for all $r>0$. The cancellation is not merely asymptotic:
the action already vanishes before the cutoff is removed. This is
Eq.~\eqref{eq:harrison-action-invariance} for the Minkowski seed, whose
action with $\sqrt{2\hat{\cR}}$ vanishes at every radius.

The expansions of the separate contributions,
Eq.~\eqref{eq:melvin-sph-bulk-GH-asymptotics}, are obtained after
integrating and are consistent with this exact identity.
The zero-field limit is different from the radial limit:
for $\ell\to\infty$ at fixed $r$,
$I^E_{\mathrm{bulk}}\to0$, $I^E_{\mathrm{GH}}\to-\beta r$, and
$I^E_{\mathrm{ct}}\to\beta r$, as appropriate for flat space.

\subsection{Schwarzschild--Melvin: regulated action and radial limit}
\label{RG-esfericas}

For $m>0$, we regulate the exterior by
$2m\leq\bar r\leq r$, with $r>2m$ and $f(r)=1-2m/r>0$.
The Euclidean metric~\eqref{eq:schmel-sph-E-metric} and its
behavior near the horizon,
Eq.~\eqref{eq:schmel-sph-bolt-regularity}, fix $\beta=8\pi m$.
The horizon is the set where the Euclidean circle
shrinks, not an inner boundary to which a
Gibbons--Hawking--York term should be added~\cite{GibbonsHawking,HawkingHorowitz}.
The outer cutoff retains the Dirichlet data established
in Sec.~\ref{sec:variational-principle}.

The bulk contribution can be evaluated by means of the
identity~\eqref{eq:schmel-sph-Maxwell-divergence} and the flux
\eqref{eq:schmel-sph-Maxwell-flux}. This conversion uses
the Maxwell equations on the solution; it neither adds a
Legendre term nor modifies the prescribed data.
The integrated expressions~\eqref{eq:schmel-sph-bulk-exact}
and~\eqref{eq:schmel-sph-GH-exact} contain nonpolynomial
terms that cancel exactly. Their sum is
\begin{equation}
  I^E_{\mathrm{bulk}}(r)+I^E_{\mathrm{GH}}(r)
       =-\beta r+\frac{3\beta m}{2}.
  \label{eq:schmel-sph-bare-exact}
\end{equation}
This is the value of the Schwarzschild seed at the same radius, as
required by the invariance of
Eq.~\eqref{eq:app-harrison-action-with-boundary}.

We apply the same counterterm,
\begin{equation}
I^E_{\mathrm{ct}}
=\frac{1}{\kappa}\int_{\Sigma}\dd^3y\,\sqrt{h_E}\,
\sqrt{2\left(\cR-\mathcal F^2\right)} .
\label{eq:schmel-sph-counterterm}
\end{equation}
The form and the normalization of this functional follow from
Eq.~\eqref{eq:harrison-counterterm-transfer}. The analysis of the
family of coefficients in
Appendix~\ref{app:prd-esfera-pura} constitutes an additional
check, not a fit to the final value of the action.

The invariants~\eqref{eq:schmel-sph-intrinsic-invariants}
and $\sqrt{h_E}$ from Eq.~\eqref{eq:schmel-sph-boundary-geometry} give
\begin{equation}
  U=\sqrt{2(\cR-\mathcal F^2)}=\frac{2}{r\Lambda},
  \qquad \sqrt{h_E}\,U=2r\sqrt f\,\sin\theta.
  \label{eq:schmel-sph-mixed-density}
\end{equation}
The simplification of $\cR-\mathcal F^2$ occurs before
integrating and over the whole sphere, including the polar annuli.
It does not mean that the electromagnetic sector of the
bulk disappears. The integral and its expansion at fixed $m$ are
\begin{equation}
\begin{aligned}
I^E_{\mathrm{ct}}(r)
&=\beta r\sqrt{f(r)}\\
&=\beta r-\beta m-\frac{\beta m^2}{2r}
-\frac{\beta m^3}{2r^2}
+\Order\!\left(\frac{\beta m^4}{r^3}\right).
\end{aligned}
\label{eq:schmel-sph-counterterm-exact}
\end{equation}
Adding this term to Eq.~\eqref{eq:schmel-sph-bare-exact},
we obtain the action at finite cutoff and its limit:
\begin{align}
I^E_{\mathrm{ren}}(r)
&=\beta\left[r\bigl(\sqrt{f(r)}-1\bigr)+\frac{3m}{2}\right],
\label{eq:schmel-sph-renormalized-exact}\\
I^{E,\infty}_{\mathrm{ren}}
&=\lim_{r\to\infty}I^E_{\mathrm{ren}}(r)=\frac{\beta m}{2}.
\label{eq:schmel-sph-renormalized-infinity}
\end{align}
The limit is taken at fixed $m$ and $\ell$; Euclidean
regularity then keeps $\beta=8\pi m$ fixed.
Unlike in Melvin, the regulated action depends on $r$:
$\beta m/2$ is its asymptotic value, not the value at every cutoff.

This value of the action was already obtained through subtraction
of the Melvin background~\cite{Radu:2002}, and its independence of the
external field in the static family was discussed
in~\cite{Ortaggio:2005}. Subtraction of the Melvin background had
been used earlier in the Euclidean actions of the instantons
that describe the pair creation of black holes in a magnetic
field~\cite{DowkerEtAl:1994,HawkingHorowitzRoss:1995}. Here it is recovered with the functional
inherited from the seed through the Harrison transformation, without background
subtraction or adjustment of coefficients. By
Eq.~\eqref{eq:harrison-action-invariance}, this agreement is a
consequence of the construction and not an independent test of it:
Eq.~\eqref{eq:schmel-sph-renormalized-exact} is the Schwarzschild action
with the counterterm $\sqrt{2\hat{\cR}}$. The independent content lies
in the boundary tensor, the current, and the Killing-time charge, which
the invariance does not determine.

The value of the action on a solution does not determine the
Brown--York tensor. Its evaluation requires the
complete first variation presented in
Sec.~\ref{sec:variational-principle}; in particular, the
vanishing of the tensor on Melvin is checked separately
in Eq.~\eqref{eq:melvin-sph-BY-ren-components}.

\section{Brown--York tensor and quasilocal energy}
\label{sec:brown-york-energy}

We evaluate the responses of the complete action obtained in
Sec.~\ref{sec:variational-principle}, keeping the outward
normal and the time normalization of
Sec.~\ref{sec:melvin-universe}. The independent induced
fields are varied first, and each solution is substituted afterwards.
The vanishing of the value of the action on Melvin does not replace
this variational evaluation~\cite{BrownYork}.

We distinguish the local components of the tensor, the proper
energy of the static observers, and the Killing-time charge. The angle-dependent lapse prevents identifying the
last two through a constant factor, even though both
vanish on Melvin.

\subsection{Melvin: vanishing responses at finite cutoff}
\label{BY-spherical-Melvin}

We consider the complete Lorentzian boundary
$\Sigma\simeq\mathbb R_t\times S^2$, located at $\bar r=r>0$.
The induced data are those of Eq.~\eqref{eq:melvin-sph-BY-induced}.
We apply Eq.~\eqref{eq:melvin-sph-BY-responses}, with the coefficient
\eqref{eq:melvin-sph-BY-ct-coefficient}, after varying the
functional. The components of the extrinsic curvature
\eqref{eq:melvin-sph-BY-K-components}, the derivatives of $\Xi$
in Eq.~\eqref{eq:melvin-sph-BY-Xi-derivatives}, and the complete Hessian
lead to the explicit cancellation
\eqref{eq:melvin-sph-BY-component-cancellation}.
The off-diagonal components vanish separately. Therefore,
\begin{equation}
  \tau^{\mathrm{ren}}_{tt}
  =\tau^{\mathrm{ren}}_{\theta\theta}
  =\tau^{\mathrm{ren}}_{\varphi\varphi}=0,
  \qquad \tau^{\mathrm{ren}}_{ab}=0.
  \label{eq:melvin-sph-BY-ren-components}
\end{equation}
The cancellation is exact for all $r>0$, not merely asymptotic.
In the static orthonormal frame
\begin{equation*}
  e_{\hat t}=\Lambda^{-1}\partial_t,\qquad
  e_{\hat\theta}=(r\Lambda)^{-1}\partial_\theta,\qquad
  e_{\hat\varphi}=\frac{\Lambda}{r\sin\theta}\partial_\varphi,
\end{equation*}
the components also vanish. The tensorial statement extends
by regularity to the poles, where the angular coordinates are
singular. No expansion at fixed angle, which could
be nonuniform near the axes, is needed.

This establishes, on this background, the analog of the cancellation of the
total renormalized tensor on the wall $\mathbb R_t\times S^2_r$
of Minkowski with the counterterm $\sqrt{2\cR}$.

\paragraph{Electromagnetic response and variational scope.}
The electromagnetic contributions of
Eq.~\eqref{eq:melvin-sph-BY-Maxwell-cancellation} give
\begin{equation}
  \mathcal J^\varphi_{\mathrm{ren}}
     =-\frac4{\kappa\ell r\Lambda}
       +\frac4{\kappa\ell r\Lambda}=0,
  \qquad \mathcal J^t_{\mathrm{ren}}
       =\mathcal J^\theta_{\mathrm{ren}}=0.
  \label{eq:melvin-sph-BY-current-zero}
\end{equation}
This vanishing response does not mean that the magnetic field or
$n_\mu F^{\mu a}$ vanish: it expresses the cancellation between the
bulk and counterterm responses.

From Eq.~\eqref{eq:melvin-sph-BY-total-variation} there follows a special
property of this background: at fixed cutoff, its first variation vanishes
even for smooth variations of the induced data, with compact temporal
support, regularity at the origin and on the axis, and within the
domain~\eqref{eq:melvin-sph-E-variation-domain} of the functional. This result follows from
$\tau^{\mathrm{ren}}_{ab}=0$ and $\mathcal J^a_{\mathrm{ren}}=0$, not from
the mere equality $I^E_{\mathrm{ren}}(r)=0$ on the solution. A variational problem for a general class
of asymptotically Melvin geometries would require, in addition, defining
the admissible perturbations, their falloff conditions, and the control
of the square root as the cutoff is removed.

\paragraph{Energy and Killing-time charge of the complete boundary.}
The section $\cB_t=\Sigma\cap\{t=\mathrm{const.}\}\simeq S^2$ is the
complete outer spatial boundary; it has no edge circles
and no corner contributions. Here $S^2$ indicates its topology; the
spatial metric is not round. Its foliation has
\begin{equation}
\begin{aligned}
  N&=\Lambda(\theta),\qquad V^i=0,\qquad
       u^a=\Lambda^{-1}(\partial_t)^a,\\
  \sigma_{ij}&=\operatorname{diag}\!\left(
           \Lambda^2r^2,\frac{r^2\sin^2\theta}{\Lambda^2}\right),
      \qquad \sqrt{\sigma}=r^2\sin\theta,
      \qquad x^i=(\theta,\varphi).
\end{aligned}
  \label{eq:melvin-sph-BY-ADM-data}
\end{equation}
We define the energy and momentum densities, the proper energy, and the
charge associated with a tangential Killing vector by
\begin{equation}
\begin{aligned}
  \varepsilon&=u^au^b\tau^{\mathrm{ren}}_{ab},\qquad
  j_i=-\sigma_i{}^a u^b\tau^{\mathrm{ren}}_{ab},\qquad
  E_{\mathrm{prop}}(r)=\int_{\cB_t}\dd^2x\,\sqrt{\sigma}\,\varepsilon,\\
  Q[\xi](r)&=\int_{\cB_t}\dd^2x\,\sqrt{\sigma}\,
                              u^a\tau^{\mathrm{ren}}_{ab}\xi^b,\\
  E_{\mathrm{BY}}[\partial_t](r)
     &=\int_{\cB_t}\dd^2x\,\sqrt{\sigma}\,
                           (N\varepsilon-V^ij_i)
       =Q[\partial_t](r).
\end{aligned}
  \label{eq:melvin-sph-BY-energy-definitions}
\end{equation}
This convention for $Q$ agrees with the sign of the Brown--York
boundary Hamiltonian and with that of the charges of Balasubramanian and
Kraus~\cite{BrownYork,BalasubramanianKraus:1999}.\footnote{Equations~(12)--(14)
of Balasubramanian and Kraus use the positive
convention adopted in Eq.~\eqref{eq:melvin-sph-BY-energy-definitions}.}
The vector $\xi=\partial_t$ is normalized by unit proper
time on the Melvin axis, where $\Lambda=1$; it is not of unit norm over
the whole of $\Sigma$. Since $N$ depends on $\theta$, it cannot be pulled out
of the integral to convert $E_{\mathrm{prop}}$ into
$E_{\mathrm{BY}}$ through a constant factor. The difference between
proper time and coordinate time remains relevant even though
both energies vanish on this background.

The vanishing of the tensor implies $\varepsilon=0$, $j_i=0$, and
\begin{equation}
  E_{\mathrm{prop}}(r)=E_{\mathrm{BY}}[\partial_t](r)
       =Q[\partial_t](r)=0\qquad(r>0).
  \label{eq:melvin-sph-BY-energy-zero}
\end{equation}
Conservation is described by the boundary identities
\eqref{eq:melvin-sph-BY-Ward}, presented in
Sec.~\ref{sec:variational-principle}.
For $\xi=\partial_t$, the static symmetry and
$\xi^b\mathcal F_{ba}=0$ imply
$D_a(\tau^{\mathrm{ren}\,a}{}_{b}\xi^b)=0$.
Since $\cB_t$ is closed, there is no flux through spatial edges.
Moreover, $A_t=0$, so the time-translation generator contains no
term of electric potential times charge~\cite{HawkingHorowitz}.

\paragraph{Meaning of the vanishing energy.}
Equation~\eqref{eq:melvin-sph-BY-energy-zero} is a normalization of the
gravitational quasilocal energy, not a statement that the spacetime
carries no electromagnetic energy. The bulk invariant $F^2$ and the
magnetic energy density measured by static observers are nonzero, and
the Maxwell term contributes to the bulk action. What vanishes is the
total renormalized response of the boundary, in which the bulk,
Gibbons--Hawking--York, and counterterm contributions cancel, as
Eq.~\eqref{eq:melvin-sph-BY-current-zero} shows explicitly for the
current. The counterterm thus assigns zero energy to Melvin, as
$\sqrt{2\hat{\cR}}$ assigns zero energy to Minkowski, and Melvin is the
zero of energy of the family with fixed $B_0$. In background
subtraction the same normalization is imposed by choosing Melvin as the
reference~\cite{Radu:2002}; here it follows from the counterterm.

\subsection{Schwarzschild--Melvin: physical tensor and Killing-time charge}
\label{BY-Sch}

For $m>0$, $\ell>0$, and $r>2m$, we evaluate the same responses
\eqref{eq:melvin-sph-BY-responses} on Schwarzschild--Melvin.
The variation with respect to independent sources precedes the
evaluation on the solution. The cancellations
\eqref{eq:schmel-sph-BY-temporal-cancellation} and
\eqref{eq:schmel-sph-BY-spatial-counterterm} include the
electromagnetic contribution and the complete Hessian.
The exact coordinate components are presented in
Eqs.~\eqref{eq:schmel-sph-BY-temporal-component},
\eqref{eq:schmel-sph-BY-polar-component}, and
\eqref{eq:schmel-sph-BY-azimuthal-component}.

\paragraph{Physical components and uniform control.}
The coordinate component~\eqref{eq:schmel-sph-BY-polar-component}
grows with $r$; as shown next, this growth reflects the coordinate
basis and not a divergent physical stress. In the
orthonormal frame of the static observers,
\begin{equation}
  e_{\hat t}=\frac{\partial_t}{\Lambda\sqrt f},\qquad
  e_{\hat\theta}=\frac{\partial_\theta}{r\Lambda},\qquad
  e_{\hat\varphi}=\frac{\Lambda}{r\sin\theta}\partial_\varphi,
  \label{eq:schmel-sph-BY-frame}
\end{equation}
the energy density and the principal stresses are
\begin{equation}
\begin{aligned}
  \varepsilon=\tau^{\mathrm{ren}}_{\hat t\hat t}
    &=\frac{2(1-\sqrt f)}{\kappa r\Lambda},\\
  p_\theta=\tau^{\mathrm{ren}}_{\hat\theta\hat\theta}
    &=\frac{(1-\sqrt f)^2}{2\kappa r\Lambda\sqrt f},\\
  p_\varphi=\tau^{\mathrm{ren}}_{\hat\varphi\hat\varphi}
    &=p_\theta-\frac{4(\Lambda-1)(1-\sqrt f)}{\kappa r\Lambda^2}.
\end{aligned}
  \label{eq:schmel-sph-BY-physical-components}
\end{equation}
At each fixed $r>2m$ they are smooth. At both poles, $\Lambda=1$,
$p_\theta=p_\varphi$, and $p_\varphi-p_\theta=\Order(\sin^2\theta)$.
Although the angular frame degenerates on the axis, this isotropy and the
order of vanishing of the anisotropy ensure the smooth extension
of the spatial tensor in local Cartesian charts.

There is, in addition, uniform control as the cutoff is removed. For
$r\geq r_0>2m$, let us define $y_0=\sqrt{1-2m/r_0}>0$.
The identities $1-\sqrt f=2m/[r(1+\sqrt f)]$,
$\Lambda\geq1$, and $(\Lambda-1)/\Lambda^2\leq1/4$ give
\begin{equation}
\begin{aligned}
  0\leq\varepsilon&\leq\frac{4m}{\kappa r^2},
  &0\leq p_\theta&\leq\frac{2m^2}{\kappa y_0r^3},\\
  |p_\varphi|&\leq p_\theta+\frac{2m}{\kappa r^2}.
\end{aligned}
  \label{eq:schmel-sph-BY-uniform-bounds}
\end{equation}
All the orthonormal components tend to zero uniformly
on the sphere. In this case, the apparent divergence of the polar
covariant component stems from the growth of the coordinate basis.
These bounds do not apply in the limit $r\to 2m^{+}$, since
the boundary becomes null and the orthonormal frame associated
with the static observers ceases to be well defined.

\paragraph{Electromagnetic response and conservation.}
The contributions of Eq.~\eqref{eq:schmel-sph-BY-Maxwell-pieces}
give the response
\begin{equation}
  \mathcal J^\varphi_{\mathrm{ren}}
     =\frac{4(1-\sqrt f)}{\kappa\ell r\Lambda},
  \qquad
  \mathcal J^t_{\mathrm{ren}}=\mathcal J^\theta_{\mathrm{ren}}=0.
  \label{eq:schmel-sph-BY-current}
\end{equation}
Unlike in Melvin, this response does not vanish for $m>0$.
Its physical component is
$\mathcal J^{\hat\varphi}_{\mathrm{ren}}
=4(1-\sqrt f)\sin\theta/(\kappa\ell\Lambda^2)$ and vanishes
regularly at the poles. It is an azimuthal response to the
magnetic potential, not a net electric charge density. Both this
current and the anisotropy $p_\varphi-p_\theta$ of
Eq.~\eqref{eq:schmel-sph-BY-physical-components} vanish if either $m$
or $B_0$ vanishes, and at leading order
$\mathcal J^\varphi_{\mathrm{ren}}=2mB_0/(\kappa r^2\Lambda)\,[1+\Order(m/r)]$.
They are local responses of the black hole to the external field, which
the integrated charge~\eqref{eq:schmel-sph-BY-charge-exact} does not
detect.

The general identities are those of Eq.~\eqref{eq:melvin-sph-BY-Ward}.
Their polar component is checked explicitly in
Eq.~\eqref{eq:schmel-sph-BY-Ward-polar}.
The metric tensor exchanges momentum with the magnetic field:
its spatial divergence need not vanish separately.
For the timelike Killing vector $\xi=\partial_t$, by contrast,
$\xi^b\mathcal F_{ba}=0$ and the symmetry of the tensor imply
$D_a(\tau^{\mathrm{ren}\,a}{}_{b}\xi^b)=0$. The Killing-time charge
is conserved on each fixed boundary. Moreover, $A_t=0$, so that
the time-translation generator requires no term of electric potential
times charge~\cite{HawkingHorowitz}.

\paragraph{Proper energy and Killing-time charge.}
On the spatial section $\cB_t=\Sigma\cap\{t=\mathrm{const.}\}$,
\begin{equation}
N=\Lambda\sqrt{f},
\qquad
u^a=\frac{1}{\Lambda\sqrt{f}}\delta^a_t,
\qquad
\sqrt{\sigma}=r^2\sin\theta .
\label{eq:schmel-sph-BY-ADM-data}
\end{equation}
With $H(r)$ defined in Eq.~\eqref{eq:schmel-sph-H}, the proper energy
and the energy associated with $\partial_t$ are distinguished
by the lapse, in accordance with Eq.~\eqref{eq:melvin-sph-BY-energy-definitions}
and the Brown--York procedure~\cite{BrownYork}:
\begin{equation}
  E_{\mathrm{prop}}(r)=\int_{\cB_t}\dd^2x\,\sqrt\sigma\,\varepsilon
       =(1-\sqrt f)H(r).
  \label{eq:schmel-sph-BY-proper-energy}
\end{equation}
By contrast,
\begin{equation}
E_{\mathrm{BY}}=\int_{\cB_t}\dd^2x\,\sqrt{\sigma}\,N\,\varepsilon,
\qquad
\varepsilon=u^a u^b\tau^{\mathrm{ren}}_{ab},
\end{equation}
and its integrated density simplifies exactly to
\begin{equation}
  \sqrt\sigma\,N\varepsilon
      =\frac{2r\sqrt f(1-\sqrt f)}\kappa\sin\theta.
  \label{eq:schmel-sph-BY-charge-density}
\end{equation}
The magnetic factor $\Lambda$ disappears before integrating, without
using any angular expansion. With $\kappa=8\pi$ we obtain
\begin{equation}
E_{\mathrm{BY}}(r)=r\left(\sqrt{f(r)}-f(r)\right).
\label{eq:schmel-sph-BY-energy-exact}
\end{equation}
This is the Killing-time charge of the Schwarzschild seed computed with
$\sqrt{2\hat{\cR}}$. Unlike the equality of the actions, it is not
implied by Eq.~\eqref{eq:harrison-action-invariance} and results from
the complete variation of Sec.~\ref{sec:variational-principle}.
The asymptotic limit is
\begin{equation}
E^{\infty}_{\mathrm{BY}}=\lim_{r\to\infty}E_{\mathrm{BY}}(r)=m.
\label{eq:schmel-sph-BY-energy-infinity}
\end{equation}
For fixed finite $\ell$, by contrast,
$E_{\mathrm{prop}}(r)=\Order(m\ell^2\ln(r/\ell)/r^2)\to0$.
This does not contradict the mass $m$: the two integrals correspond to
different time generators. The vector $\partial_t$ is
normalized by $-\xi^2\to1$ on the asymptotic axis, but it is not
of unit norm over the whole boundary. Since $N$ depends on $\theta$, it
cannot be pulled out of the integral as a uniform redshift
factor. A decaying local density can produce a finite
charge when combined with the appropriate area and lapse.

Equivalently, the conserved charge associated with the timelike Killing vector is
\begin{equation}
Q[\xi](r)
=\int_{\cB_t}\dd^2x\,\sqrt{\sigma}\,
u^a\tau^{\mathrm{ren}}_{ab}\xi^b
=r\left(\sqrt{f(r)}-f(r)\right),
\qquad
\xi^a=(\partial_t)^a=\delta^a_t .
\label{eq:schmel-sph-BY-charge-exact}
\end{equation}

The above conservation is conservation in time at fixed cutoff, not independence
of the regulator radius. Indeed,
\begin{equation}
  \frac{\dd Q[\partial_t](r)}{\dd r}
        =\frac{(1-\sqrt f)^2}{2\sqrt f}>0,
  \qquad
  Q[\partial_t](r)=m-\frac{m^2}{2r}+\Order\!\left(\frac{m^3}{r^2}\right).
  \label{eq:schmel-sph-BY-radial-dependence}
\end{equation}
Its limit defines the mass for
the specified time normalization and external field.

As checks, when $m\to0$ the tensor,
the current, and both energies vanish exactly, recovering Melvin. If
$\ell\to\infty$ at fixed $r>2m$, $H(r)\to r$,
$p_\varphi-p_\theta\to0$, and the current disappears: the
Schwarzschild proper energy $r(1-\sqrt f)$ is recovered, whereas
its charge for the coordinate time is $r\sqrt f(1-\sqrt f)$.
Finally, the action and the charge, computed independently,
satisfy the exact relation
\eqref{eq:schmel-thermo-finite-statistical-relation},
$I^E_{\mathrm{ren}}(r)=\beta Q[\partial_t](r)-\cA_H/4$,
with $\beta=8\pi m$ and $\cA_H=16\pi m^2$. Its
thermodynamic interpretation is developed in
Sec.~\ref{subsec:schmel-thermodynamic-consistency}.

\section{Thermodynamic consistency at fixed external field}
\label{subsec:schmel-thermodynamic-consistency}

The temperature, the horizon area, and the Euclidean action of
Schwarzschild--Melvin are known~\cite{Radu:2002,Ortaggio:2005}, and by
Eq.~\eqref{eq:harrison-action-invariance} the regulated action
coincides with that of the Schwarzschild seed. The check carried out
here is different: it relates the action of
Sec.~\ref{sec:euclidean-action} to the Killing-time charge of
Sec.~\ref{sec:brown-york-energy}, which follows from the complete
variation and is not fixed by the invariance of the action. We consider
the magnetostatic family with fixed external field $B_0=2/\ell$,
azimuthal period $2\pi$, and generator $\xi=\partial_t$, normalized by
$-\xi^2\to1$ on the asymptotic axis. The potential satisfies
$A_\mu\xi^\mu=0$, and the black hole carries neither electric charge
nor monopole magnetic charge. The period $\beta$ refers to this Killing
time. A static observer on the boundary measures the local temperature
$T_{\mathrm H}/N$, with $N=\sqrt{-\xi^2}$ and
$T_{\mathrm H}=\beta^{-1}$~\cite{Tolman:1930}, so that
$T_{\mathrm H}$ is not a uniform proper temperature over the regulator.

\subsection{Statistical relation at finite cutoff}

The regularity of the nonextremal Euclidean horizon,
Eq.~\eqref{eq:schmel-sph-bolt-regularity}, and the cancellation of the
magnetic factors in its area element give
\begin{equation}
\beta=8\pi m,
\qquad T_{\mathrm H}=\frac1{8\pi m},
\qquad \cA_{\mathrm H}=16\pi m^2,
\qquad m>0.
\label{eq:schmel-thermo-horizon-data}
\end{equation}
The exact expressions~\eqref{eq:schmel-sph-renormalized-exact}
and~\eqref{eq:schmel-sph-BY-charge-exact}, computed independently,
satisfy
\begin{equation}
I^E_{\mathrm{ren}}(r)
=\beta Q[\partial_t](r)-\frac{\cA_{\mathrm H}}4,
\qquad \beta=8\pi m,
\label{eq:schmel-thermo-finite-statistical-relation}
\end{equation}
before the regulator is removed, in agreement with the Hamiltonian
relation between the Euclidean action, the energy, and the area of a
nonextremal horizon~\cite{GibbonsHawking,HawkingHorowitz}. The entropy
is therefore
\begin{equation}
S=\beta Q[\partial_t](r)-I^E_{\mathrm{ren}}(r)
=\frac{\cA_{\mathrm H}}4=4\pi m^2.
\label{eq:schmel-thermo-entropy}
\end{equation}
The area term is the contribution of the horizon, where the Euclidean
circle shrinks; no inner boundary is introduced there. The identity
does not imply $Q[\partial_t](r)=(\partial_\beta I^E_{\mathrm{ren}})_{r,\ell}$:
varying $m$ at fixed coordinates of the regulator also changes the
induced metric and the lapse, and a thermodynamic problem in a finite
cavity requires the corresponding work
terms~\cite{BrownYork,York:1986,BradenBrownWhitingYork}. The
derivatives below refer to the asymptotic action.

\subsection{Asymptotic quantities and first law}

On the regular family, $m=\beta/(8\pi)$, and the asymptotic values are
\begin{align}
I^{E,\infty}_{\mathrm{ren}}
&\equiv\lim_{r\to\infty}I^E_{\mathrm{ren}}(r)
=\frac{\beta m}{2}=\frac{\beta^2}{16\pi},
\label{eq:schmel-thermo-action-infinity}\\
Q_\infty[\partial_t]
&\equiv\lim_{r\to\infty}Q[\partial_t](r)=m.
\label{eq:schmel-thermo-charge-infinity}
\end{align}
The semiclassical relations between the action and the partition
function~\cite{GibbonsHawking,HawkingHorowitz} give
\begin{equation}
U_{\mathrm{th}}
=\left(\frac{\partial I^{E,\infty}_{\mathrm{ren}}}{\partial\beta}\right)_{\ell}
=m=Q_\infty[\partial_t],
\qquad
S=\beta U_{\mathrm{th}}-I^{E,\infty}_{\mathrm{ren}}
=\frac{\cA_{\mathrm H}}4.
\label{eq:schmel-thermo-internal-energy}
\end{equation}
The derivative compares regular solutions with different $m$ at fixed
$\ell$; it is not a variation at fixed period and does not replace the
Dirichlet problem of Sec.~\ref{sec:variational-principle}. With
$\dd B_0=0$,
\begin{equation}
\dd U_{\mathrm{th}}=\dd m
=T_{\mathrm H}\,\dd S,
\qquad
U_{\mathrm{th}}=2T_{\mathrm H}S.
\label{eq:schmel-thermo-first-law-Smarr}
\end{equation}
The second equality, the Smarr relation in this sector, follows from
the computed quantities without postulating a scaling symmetry at fixed
$\ell$. The heat capacity $C_{B_0}=(\partial U_{\mathrm{th}}/\partial
T_{\mathrm H})_{B_0}=-8\pi m^2<0$ shows that this consistency does not
imply canonical thermal stability~\cite{York:1986}.

\paragraph{Scope at fixed external field.}
The first law~\eqref{eq:schmel-thermo-first-law-Smarr} does not establish
a universal first law without magnetic work.
For more general magnetized black holes, Gibbons, Pang, and Pope
obtained a term $-\mu\,\dd B$ when the external field is allowed to vary
\cite{GibbonsPangPope:2014}. Astorino and collaborators distinguished the
choice of an integrable mass when all the parameters are varied from the
alternative prescriptions that treat the magnetic field as an
external source~\cite{AstorinoEtAl:2016}. In the family studied here, the
responses of Sec.~\ref{sec:brown-york-energy} allow a direct check of
the term conjugate to $B_0$.

\paragraph{Response to the external field.}
Changing $B_0=2/\ell$ at fixed cutoff and fixed $m$, hence at fixed
$\beta$ and with a regular horizon, changes the boundary data through
\begin{equation}
\partial_{B_0}\Lambda=\frac{G}{\ell},\qquad
\partial_{B_0}A_\varphi=\frac{G(2-\Lambda)}{2\Lambda^2},\qquad
G=r^2\sin^2\theta .
\label{eq:schmel-thermo-B-variation}
\end{equation}
With the components~\eqref{eq:schmel-sph-BY-physical-components} and
the current~\eqref{eq:schmel-sph-BY-current}, the two terms of the first
variation~\eqref{eq:melvin-sph-BY-total-variation} are
\begin{equation}
\sqrt{-h}\,\mathcal J^\varphi_{\mathrm{ren}}\,\partial_{B_0}A_\varphi
=\frac{2\sqrt f\,(1-\sqrt f)\,rG(2-\Lambda)}{\kappa\ell\Lambda^2}\,\sin\theta
=-\frac12\sqrt{-h}\,\tau^{\mathrm{ren}\,ab}\,\partial_{B_0}h_{ab}.
\label{eq:schmel-thermo-B-response}
\end{equation}
The current and metric contributions cancel pointwise on the boundary,
although each is nonzero. The response conjugate to $B_0$ therefore
vanishes at every cutoff, in agreement with the independence of
Eq.~\eqref{eq:schmel-sph-renormalized-exact} of $\ell$, and
asymptotically
$\mu\equiv-\beta^{-1}(\partial I^{E,\infty}_{\mathrm{ren}}/\partial B_0)_\beta=0$.
A first law with $\dd B_0\neq0$ thus reduces, in this family, to
Eq.~\eqref{eq:schmel-thermo-first-law-Smarr}. This agrees with
Ref.~\cite{GibbonsPangPope:2014}, where the magnetic moment of
magnetized Kerr--Newman black holes is, at linear order in the electric
charge, proportional to that charge [Eq.~(5.16) there] and vanishes for
an uncharged black hole. Testing the construction on families with
$\mu\neq0$, such as Kerr--Newman--Melvin, requires the enlarged
reduction discussed in Sec.~\ref{app:counterterm-checks} and is left
for future work.

\section{Discussion and conclusions}
\label{app:counterterm-checks}

We have constructed a boundary counterterm that renormalizes the
Einstein--Maxwell action of Melvin and Schwarzschild--Melvin on
complete constant-radius boundaries, without background subtraction,
and we have derived from it the Brown--York tensor, the boundary
current, and the Killing-time charge. The values of the action and of
the mass are not new. What is new is the following. First, the
counterterm $\sqrt{2\cR}$ of asymptotically flat spacetimes leaves a
logarithmic divergence on these geometries. Second, the counterterm
$\sqrt{2(\cR-\mathcal F^2)}$ that replaces it is not an ansatz: it is
the image of $\sqrt{2\cR}$ under the Harrison transformation, which
therefore acts on the renormalized action and not only on the
solutions. Third, the quasilocal quantities follow from a specified set
of boundary data and from the complete variation of the action, rather
than from a reference background whose matching with the solution must
be chosen.

The counterterm $\sqrt{2\cR}$, used for boundaries of asymptotically
flat spacetimes~\cite{Mann:1999,KrausLarsenSiebelink,AstefaneseiMannStelea:2007},
leaves a logarithmic divergence. At fixed $\ell>0$ and $m$, without
additional subtractions, Eqs.~\eqref{eq:melvin-sph-pure-log} and
\eqref{eq:schmel-sph-pure-action-log} give
\begin{equation}
I^E_{\mathrm g}(r)
=\beta\ell\left[\ln\frac{4r}{\ell}-1\right]
+\begin{cases}
0,&\text{Melvin},\\[2pt]
\beta m/2,&\text{Schwarzschild--Melvin},
\end{cases}
+o(1),
\label{eq:conclusion-geometric-divergence}
\end{equation}
where $I^E_{\mathrm g}=I^E_{\mathrm{bulk}}+I^E_{\mathrm{GH}}
+I^E_{\mathrm{ct,g}}$. The coefficient $\beta\ell$ is the same in both
solutions, and the divergence originates in the polar annuli
$\ell/r\ll\vartheta\ll\sqrt{\ell/r}$, with
$\vartheta=\min(\theta,\pi-\theta)$, where the expansion at fixed angle
fails; the exact integral and the bound on the remainder of
Appendix~\ref{app:prd-esfera-pura} establish it. This result is in
contrast with the statement of Ref.~\cite{Radu:2002} that the same
functional yields a finite action: a finite value is obtained if only
the leading term of the fixed-angle expansion is integrated. The value
$\beta m/2$ obtained there by background subtraction is unaffected.

The intrinsic identity~\eqref{eq:intro-transport} follows from the
axial reduction and holds pointwise, without field equations, for all
boundary data in the hypersurface-orthogonal axial sector,
Eq.~\eqref{eq:app-harrison-general-density-identity};
Appendix~\ref{app:harrison-charged-seeds} checks it explicitly for a
magnetized seed and for a magnetically charged seed. It fixes the
electromagnetic term and its coefficient without fitting them to the
mass or to an integrated cancellation, and the resulting counterterm is
form invariant along the Harrison orbit. Combined with the invariance
of the bulk and Gibbons--Hawking--York terms, with the treatment of the
axis of Appendix~\ref{app:harrison-GHY}, it implies that the
renormalized action of each magnetized solution equals that of its seed
at every cutoff, Eq.~\eqref{eq:harrison-action-invariance}. A seed in
this sector renormalized by $\sqrt{2\hat{\cR}}$ on a given regulator is
thus mapped to a magnetized solution renormalized by
$\sqrt{2(\cR-\mathcal F^2)}$, with the same value of the action at each
cutoff: the domain of validity of the new counterterm is the Harrison
image of that of the seed counterterm. This explains the independence of the action of the external field noted
in~\cite{Radu:2002,Ortaggio:2005}, and it means that the agreement with
background subtraction is a consequence of the construction rather than
an independent test of it.

The independent content lies in the responses. Since the counterterm
is intrinsic, the Dirichlet problem with fixed induced metric and
tangential potential remains stationary at finite cutoff, and the
complete variation of Sec.~\ref{sec:variational-principle} yields
$\tau^{\mathrm{ren}}_{ab}$ and $\mathcal J^a_{\mathrm{ren}}$, which
satisfy the Ward identities~\eqref{eq:melvin-sph-BY-Ward}. Obtaining
them requires specifying the data held fixed, the domain $X>0$ of the
functional, and the regularity on the axis; the responses are then
defined by the boundary data of the solution itself rather than by its
matching with a reference. For Melvin, the tensor, the current, and the
action vanish exactly at every cutoff. This is a normalization of the
gravitational energy, which takes Melvin as the zero of the family with
fixed $B_0$, and not a statement that the electromagnetic field carries
no energy. For Schwarzschild--Melvin the regulated
action and charge depend on the cutoff radius, and their limits are
\begin{equation}
\bigl(I^{E,\infty}_{\mathrm{ren}},Q_\infty[\partial_t]\bigr)
=\begin{cases}
(0,0),&\text{Melvin},\\[2pt]
(\beta m/2,m),&\text{Schwarzschild--Melvin},\quad\beta=8\pi m.
\end{cases}
\label{eq:conclusion-asymptotic-results}
\end{equation}
The charge $m$ is that of $\partial_t$, normalized on the asymptotic
axis; its integral includes the lapse $N=\Lambda\sqrt f$, and its
independence of $\ell$ stems from an exact cancellation in its density,
even though the local responses depend on the magnetic field. At finite
cutoff the action and the charge, computed independently, satisfy
$I^E_{\mathrm{ren}}(r)=\beta Q[\partial_t](r)-\cA_{\mathrm H}/4$, and in
the asymptotic sector at fixed $B_0$ they satisfy the first law and the
Smarr relation with $S=\cA_{\mathrm H}/4$.

The explicit evaluations are restricted to the two exact families, the
complete spherical regulator, and Dirichlet conditions at finite
cutoff, and the uniqueness of the counterterm is not established. Three
extensions remain. First, an asymptotic variational principle for
general perturbations requires falloff conditions and control of the
responses, since $X=2/(r^2\Lambda^2)$ tends to zero as the cutoff is
removed. Second, in the static family
the response conjugate to $B_0$ vanishes pointwise
(Sec.~\ref{subsec:schmel-thermodynamic-consistency}), in agreement with
Ref.~\cite{GibbonsPangPope:2014}; families with a nonzero magnetic
moment, such as Kerr--Newman--Melvin, would test the corresponding
magnetic work terms and the integrability of the
charges~\cite{GibbonsPangPope:2014,AstorinoEtAl:2016}. Third, the
identity covers the whole hypersurface-orthogonal axial sector,
including seeds with nonvanishing potential, but its global use
requires a regular gauge on the axis: for a magnetically charged seed
the potential cannot vanish at both poles, and the image has conical
singularities except possibly in a particular gauge
(Appendix~\ref{app:harrison-charged-seeds}). Rotating solutions such as
Kerr--Melvin and
Kerr--Newman--Melvin~\cite{ErnstWild:1976,GibbonsMujtabaPope:2013,BoothEtAl:2015}
lie outside this sector, since $g_{i\varphi}\neq0$ and the potential
acquires components $A_i$; their treatment requires enlarging the
reduction, and the invariance of the density has not been established
for them. The application of the construction to other magnetized
solutions will be addressed in a separate work.

\section*{Data availability}
No data were created or analyzed in this study.

\begin{acknowledgments}
D.C. thanks the Pontificia Universidad Cat\'olica del Per\'u (PUCP) for its
support.

A.H. gratefully acknowledges the Pontificia Universidad
Cat\'olica del Per\'u for providing a stimulating academic
environment that has fostered his scientific training
and the development of this research.

AI assistants (Claude Fable 5.2 and Opus 5.5, Anthropic; GPT-6 Sol and
GPT-6 Astra, OpenAI) were used to help organize the material, to draft
and edit parts of the text, and to cross-check calculations, including
symbolic checks with SymPy of the examples in
Appendix~\ref{app:harrison-charged-seeds}. The authors verified all
results independently and take full responsibility for the content.
\end{acknowledgments}

\appendix

\section{Polar integrals and asymptotic control}
\label{app:prd-esfera-pura}

This appendix collects the exact representation, the bounds on the remainder,
and the analysis by angular regions that support
Sec.~\ref{sec:geometric-counterterm-limit}. At the end we examine
the dependence on the electromagnetic coefficient within
the family of functionals considered in the manuscript.

We compare the functionals $I^E_{\mathrm{ct,g}}$ and
$I^E_{\mathrm{ct}}$ of Eqs.~\eqref{eq:melvin-sph-pure-counterterm}
and~\eqref{eq:melvin-sph-counterterm} on the same complete boundary
$r=\mathrm{const.}$, with $\tau\sim\tau+\beta$ and
$\varphi\sim\varphi+2\pi$. The analysis is carried out at fixed $\ell>0$,
before taking the radial limit, without an additional background
subtraction or an angular excision of the poles. It is valid for Melvin,
$f=1$, and for Schwarzschild--Melvin, $f=1-2m/r$ with $r>2m$:
$f$ is constant on each boundary and only introduces the factor
$\sqrt f$ in $\sqrt{h_E}$.

\subsection{Exact representation and bound on the remainder}

With $x=r/\ell$ and using the symmetry under $\theta\mapsto\pi-\theta$,
the exact difference is
\begin{align}
  \Delta I^E_{\mathrm{ct}}(r)
    &\equiv I^E_{\mathrm{ct,g}}(r)-I^E_{\mathrm{ct}}(r)
      =\beta\ell\sqrt f\,D(x),
      \label{eq:polar-ct-difference}\\
  D(x)&=x\int_0^{\pi/2}\dd\theta\,\sin\theta
       \left[\sqrt{1+
       \frac{4x^2\cos^2\theta}{(1+x^2\sin^2\theta)^2}}-1\right].
      \label{eq:polar-D-angular}
\end{align}
The integrand is nonnegative and $D(x)>0$ for all $x>0$.
The transformation $u=\cos\theta$, followed by
\begin{equation}
  s=\frac{1+x^2(1-u^2)}{2xu},\qquad
  u=\frac{\sqrt{s^2+x^2+1}-s}{x},\qquad
  -x\frac{\dd u}{\dd s}
    =1-\frac{s}{\sqrt{s^2+x^2+1}},
  \label{eq:polar-variable-change}
\end{equation}
maps the pole $u=1$ to $s=1/(2x)$ and the equator $u=0$ to
$s=\infty$. Thus,
\begin{equation}
  D(x)=\int_{1/(2x)}^\infty\dd s\,
       \bigl(\sqrt{1+s^{-2}}-1\bigr)
       \left(1-\frac{s}{\sqrt{s^2+x^2+1}}\right).
  \label{eq:polar-D-exact}
\end{equation}
This representation retains the poles without introducing a nonuniform
angular expansion. We write $D(x)=L(x)-E_{\mathrm{err}}(x)$, where
\begin{align}
  L(x)&=\int_{1/(2x)}^\infty\dd s\,
                     \bigl(\sqrt{1+s^{-2}}-1\bigr),\\
  E_{\mathrm{err}}(x)&=\int_{1/(2x)}^\infty\dd s\,
       \frac{\sqrt{1+s^2}-s}{\sqrt{s^2+x^2+1}}.
  \label{eq:polar-error-integral}
\end{align}
An antiderivative of the integrand of $L$ is
\begin{equation*}
  P(s)=\sqrt{1+s^2}
       +\ln\frac{s}{1+\sqrt{1+s^2}}-s,
  \qquad P(\infty)=0.
\end{equation*}
Consequently,
\begin{equation}
  L(x)=-P\!\left(\frac1{2x}\right)
       =\ln(4x)-1+\Order(x^{-1}).
  \label{eq:polar-L-asymptotic}
\end{equation}
To control the remainder, we use
$0<\sqrt{1+s^2}-s\leq\min\{1,(2s)^{-1}\}$.
If $x\geq1$, splitting the integral into $[1/(2x),1]$, $[1,x]$,
and $[x,\infty)$ yields the bound
\begin{equation}
  0<E_{\mathrm{err}}(x)
    \leq\frac1x+\frac{\ln x}{2x}+\frac1{2x}
    =\frac{3+\ln x}{2x}.
  \label{eq:polar-error-bound}
\end{equation}
Therefore, without interchanging a nonuniform limit with the integral,
\begin{equation}
  D(x)=\ln(4x)-1+
       \Order\!\left(\frac{\ln x}{x}\right),
  \qquad x\to\infty.
  \label{eq:polar-D-asymptotic}
\end{equation}
For Melvin, $I^E_{\mathrm{bulk}}+I^E_{\mathrm{GH}}=-\beta r$ and
$I^E_{\mathrm{ct}}=\beta r$ at every cutoff. Substituting
Eq.~\eqref{eq:polar-D-asymptotic} into Eq.~\eqref{eq:polar-ct-difference}
gives Eq.~\eqref{eq:melvin-sph-pure-log}, including its coefficient
and its constant term. For Schwarzschild--Melvin the radial
factor is retained:
\begin{equation}
  \Delta I^E_{\mathrm{ct}}(r)
    =\beta\ell\sqrt{1-\frac{2m}{r}}\,
       \left[\ln\frac{4r}{\ell}-1
       +\Order\!\left(\frac{\ell}{r}\ln\frac{r}{\ell}\right)\right].
  \label{eq:polar-schmel-difference}
\end{equation}
When $m$ is kept fixed, the expansion of the square root adds
subleading terms of order $\beta m\ell\ln(r/\ell)/r$ and does not remove
the logarithm. Since changing the counterterm does not change the
bulk and Gibbons--Hawking--York terms, this same
difference separates the two regulated total actions.

\subsection{Angular localization and expansion at fixed angle}

The exact representation above determines both the logarithm
and its constant term and controls the remainder. The analysis by
angular regions makes it possible to localize the logarithmic contribution,
without replacing that control of the complete integral.

The intermediate angular region
$x^{-1}\ll\theta\ll x^{-1/2}$ helps to localize the origin of the
result: there the integrand of Eq.~\eqref{eq:polar-D-angular},
including the factor $x$, behaves as $2/\theta$.
The integral accumulates a logarithm; the opposite hemisphere is already
included through the symmetry used in Eq.~\eqref{eq:polar-ct-difference}.
The field and the geometry are regular at both poles. What fails
is the uniformity of the expansion at fixed angle, not the regularity
of the solution. The zero-field limit $\ell\to\infty$ at fixed $r$
lies outside this expansion in $r/\ell\to\infty$; at zero field
the two functionals coincide.

\paragraph{Control by angular regions.}
We use the function $g_x$ defined in Eq.~\eqref{eq:melvin-sph-polar-excess}.
The logarithmic coefficient can be controlled without extending that
approximation to the ends of the annulus. For large $x$, we divide
the hemisphere into $[0,x^{-1}]$, $[x^{-1},x^{-1/2}]$, and
$[x^{-1/2},\pi/2]$. The inequalities
$0\leq\sqrt{1+q}-1\leq\min\{\sqrt q,q/2\}$ give
\begin{equation}
  \int_0^{x^{-1}}g_x\,\dd\theta\leq\ln2,
  \qquad
  \int_{x^{-1/2}}^{\pi/2}g_x\,\dd\theta=\Order(1).
  \label{eq:melvin-sph-polar-core-outer}
\end{equation}
In the intermediate interval,
\begin{equation}
  g_x=\partial_\theta\ln\Lambda-x\sin\theta+e_x,
  \qquad
  0\leq e_x\leq\frac{\sin\theta\,\Lambda}{4\cos\theta}.
  \label{eq:melvin-sph-polar-annulus}
\end{equation}
The integrals of the last two terms are bounded uniformly
in $x$, whereas
$\ln[\Lambda(x^{-1/2})/\Lambda(x^{-1})]
=\ln x-\ln2+\Order(x^{-1})$. Therefore, $D(x)=\ln x+\Order(1)$.

\paragraph{Fixed-angle expansion of Schwarzschild--Melvin.}
We consider the density~\eqref{eq:schmel-sph-pure-density} and
use $\vartheta$ for the angular distance to the corresponding pole,
as in Sec.~\ref{sec:geometric-counterterm-limit}.
At fixed angle $0<\theta<\pi$ and $x\to\infty$,
\begin{equation}
  w_g=2r\sqrt f\,\sin\theta
    +\frac{4\ell^2\sqrt f}{r}
       \frac{\cos^2\theta}{\sin^3\theta}
    +\Order_\theta\!\left(\frac{\ell^4}{r^3}\right).
  \label{eq:schmel-sph-pure-fixed-angle}
\end{equation}
The first correction behaves as $\vartheta^{-3}$ and cannot
be integrated up to the poles. The fact that its coefficient
tends to zero does not justify interchanging the expansion and the integral.
Indeed, $q_x\to0$ at fixed angle, whereas $q_x(0)=q_x(\pi)=4x^2$.
The approximation $\Lambda\sim x^2\vartheta^2$ requires
$\vartheta\gg x^{-1}$, but expanding the square root also requires
$q_x\ll1$, that is, $\vartheta\gg x^{-1/2}$.

\subsection{Family of electromagnetic coefficients}

For the family of counterterms with density
$\sqrt{h_E}\sqrt{2(\cR-a\mathcal F^2)}$, we denote the functional
by $I^E_{\mathrm{ct},a}$ and consider fixed $q=1-a>0$.
The difference between $I^E_{\mathrm{ct},a}$ and the
mixed counterterm on Melvin is $\beta\ell D_q(x)$, where the
same transformation~\eqref{eq:polar-variable-change} gives
\begin{equation}
  D_q(x)=\int_{1/(2x)}^\infty\dd s\,
    \left(\sqrt{1+\frac q{s^2}}-1\right)
    \left(1-\frac{s}{\sqrt{s^2+x^2+1}}\right).
  \label{eq:polar-family-D-exact}
\end{equation}
An antiderivative of the first factor is
\begin{equation}
  P_q(s)=\sqrt{s^2+q}
      +\sqrt q\ln\frac{s}{\sqrt q+\sqrt{s^2+q}}-s,
  \qquad P_q(\infty)=0.
  \label{eq:polar-family-primitive}
\end{equation}
Therefore, $D_q(x)=-P_q(1/(2x))-E_q(x)$, with
\begin{equation}
  E_q(x)=\int_{1/(2x)}^\infty\dd s\,
        \frac{\sqrt{s^2+q}-s}{\sqrt{s^2+x^2+1}}.
  \label{eq:polar-family-error}
\end{equation}
The inequality
$0<\sqrt{s^2+q}-s\leq\min\{\sqrt q,q/(2s)\}$ and the splitting
into the three intervals of $s$ used to obtain
Eq.~\eqref{eq:polar-error-bound} yield, for $x\geq1$,
\begin{equation}
  0<E_q(x)\leq\frac{\sqrt q}{x}
              +\frac{q\ln x}{2x}+\frac q{2x}.
  \label{eq:polar-family-error-bound}
\end{equation}
The evaluation of $P_q$ at the lower endpoint then leads to
\begin{equation}
  D_q(x)=\sqrt q\,[\ln(4\sqrt q\,x)-1]
       +\Order_q\!\left(\frac{\ln x}{x}\right),
  \qquad x\to\infty,
  \label{eq:polar-family-D-asymptotic}
\end{equation}
which exhibits a logarithmic excess for fixed $a<1$. For $q=0$ the
exact difference vanishes identically. This last case can be read off
directly from Eq.~\eqref{eq:polar-family-D-exact}; the expansion
\eqref{eq:polar-family-D-asymptotic} has been formulated for fixed $q>0$
and its remainder depends on $q$.

Thus, within the examined values of this family, $a<1$
retains a logarithmic excess, whereas $a=1$ coincides with
the mixed counterterm. This check by means of the complete
integral does not establish uniqueness among more general functionals,
nor does it replace the Harrison-based foundation developed in
Sec.~\ref{sec:harrison-construction}.

\section{Axial reduction, boundary terms, and intrinsic identity}
\label{app:harrison-n4}

We collect the details that support the construction of
Sec.~\ref{sec:harrison-construction}: the reduction of the action,
the normalization of the magnetizing map, the treatment of the axis,
an alternative check of the intrinsic identity, and two explicit
examples with seeds of nonvanishing potential.
We work with the Harrison transformation
\cite{Harrison:1968}, whose application to Schwarzschild gives the
Ernst solution~\cite{Ernst:1976}, and with the specialization
to four dimensions of Appendix~A of~\cite{Ortaggio:2005}.
We keep the Einstein--Maxwell normalization of
Eq.~\eqref{eq:lorentzian-bulk-GH} and the divergence terms
that enter the action with boundary.

\subsection{Details of the axial reduction}
\label{app:harrison-reduction}

We start from the ansatz~\eqref{eq:app-harrison-ansatz}, with
$\eta=\partial_\varphi$ spacelike, hypersurface orthogonal,
and of period $2\pi$. In the region $V>0$, all the fields are
independent of $\varphi$ and $\bar g$ has signature $(-++)$;
$\psi=0$ is not assumed. We denote by $\bar D$, $\bar\Box$,
and $\bar R$ the connection, the wave operator, and the scalar
curvature of $\bar g$. The additional components of the
four-dimensional connection are
\begin{equation}
  \Gamma^\varphi{}_{i\varphi}=\frac{\partial_iV}{2V},
  \qquad
  \Gamma^i{}_{\varphi\varphi}=-\frac12\bar g^{ij}\partial_jV.
  \label{eq:app-harrison-connection}
\end{equation}
Their contraction gives, with $\Phi=\sqrt V$,
\begin{equation}
  \sqrt{-g}=\Phi\sqrt{-\bar g},\qquad
  R[g]=\bar R-2\Phi^{-1}\bar\Box\Phi,\qquad
  F^2=\frac2V(\bar D\psi)^2.
  \label{eq:app-harrison-curvature-reduction}
\end{equation}
Therefore, the bulk density satisfies
\begin{equation}
\begin{aligned}
  \sqrt{-g}(R-F^2)
    ={}&\sqrt{-\bar g}\sqrt V
             \left[\bar R-\frac2V(\bar D\psi)^2\right]\\
      &-2\partial_i\left(\sqrt{-\bar g}\,\bar g^{ij}
                                \partial_j\sqrt V\right).
\end{aligned}
  \label{eq:app-harrison-Jordan-density}
\end{equation}
Integrating over $\varphi$ and omitting the second line, one obtains
the reduced action of Eq.~(A.5) of~\cite{Ortaggio:2005},
with $\Delta\varphi/(16\pi)=2\pi/(2\kappa)$. Here that
divergence is retained in order to follow its contribution to the boundary.

In terms of $\gamma_{ij}=V\bar g_{ij}$ and $u=\ln(V/V_*)$,
defined in Eq.~\eqref{eq:app-harrison-Einstein-frame}, the ansatz
reads
\begin{equation}
  \dd s^2=V^{-1}\gamma_{ij}\dd x^i\dd x^j+V\dd\varphi^2,
  \qquad A=\psi\,\dd\varphi,\qquad V=g(\eta,\eta)>0.
  \label{eq:harrison-reduction-ansatz}
\end{equation}
The constant $V_*>0$ has the units of $V$. In this reduction,
$D$ and $\Box_\gamma$ refer to $\gamma$, and the unbarred
contractions are performed with it. The conformal transformation in
three dimensions gives
\begin{equation}
\begin{aligned}
  \bar R&=V\left[R[\gamma]+2\Box_\gamma u
                                     -\tfrac12(Du)^2\right],\\
  \bar\Box\sqrt V&=\tfrac12V^{3/2}\Box_\gamma u.
\end{aligned}
  \label{eq:app-harrison-conformal-reduction}
\end{equation}
Substituting these expressions into Eq.~\eqref{eq:app-harrison-Jordan-density}
produces the exact identity~\eqref{eq:app-harrison-exact-density}.
The term $\Box_\gamma u$ appears with positive sign and
unit coefficient; its contribution cancels against the
corresponding part of the Gibbons--Hawking--York term.

After integrating over the axial circle and separating that divergence, one
obtains Eq.~\eqref{eq:harrison-reduced-action}. Its equations are
\begin{equation}
\begin{aligned}
  R_{ij}[\gamma]&=\frac{\partial_iV\partial_jV}{2V^2}
                     +\frac2V\partial_i\psi\partial_j\psi,\\
  \Box_\gamma V-\frac{(DV)^2}{V}+2(D\psi)^2&=0,
  \qquad D_i\left(V^{-1}D^i\psi\right)=0.
\end{aligned}
  \label{eq:app-harrison-reduced-eom}
\end{equation}
Under the ansatz, the mixed Einstein equations with indices
$i\varphi$ and the Maxwell equations with free index $i$
vanish. The solutions of the reduced equations thus
reconstruct four-dimensional solutions in the hypersurface-orthogonal axial
and aligned electromagnetic sector considered.

\subsection{Normalization and properties of the map}
\label{app:harrison-general-map}

With $\zeta=\psi+\I\sqrt V$, the scalar sector has the metric
\eqref{eq:harrison-target-metric}. The isometry
\eqref{eq:app-harrison-isometry} shows that the numerator and the
denominator acquire the same factor $|1+b\zeta|^{-4}$.
Since $\gamma$ remains fixed, the reduced density without total
divergences is preserved and, with it,
Eqs.~\eqref{eq:app-harrison-reduced-eom}. The complete bulk
term changes according to Eq.~\eqref{eq:app-harrison-bulk-divergence};
its compensation by the Gibbons--Hawking--York term is given in
Eq.~\eqref{eq:app-harrison-action-with-boundary}.

The real map~\eqref{eq:app-harrison-general-map} corresponds
to Eq.~(A.4) of~\cite{Ortaggio:2005} for $n=4$:
$2/(n-3)=2$, $(n-3)/(n-2)=1/2$, and the parameter of that
reference is $B=2b$. The Maxwell normalization coincides
with that of our action, with no additional redefinition of $A$.

If $H_b$ denotes the map~\eqref{eq:harrison-mobius}, one has
$H_c\circ H_b=H_{b+c}$ and $H_b^{-1}=H_{-b}$.
The map does not require a weak-field expansion. The explicit
expression of the transformed field is
\begin{equation}
  \dd\psi'=\frac{[(1+b\psi)^2-b^2V]\,\dd\psi
                      +b(1+b\psi)\,\dd V}{\Lambda^2},
  \qquad F'=\dd\psi'\wedge\dd\varphi.
  \label{eq:app-harrison-field-map}
\end{equation}
These results do not include the sectors with $g_{i\varphi}\ne0$
or independent components $A_i$, whose treatment would require
enlarging the set of fields of the reduction.

\subsection{Boundary term and treatment of the axis}
\label{app:harrison-GHY}

Away from the axis, the cancellation is expressed by
Eqs.~\eqref{eq:app-harrison-GHY-reduction} and
\eqref{eq:app-harrison-action-with-boundary}. After integrating over
the axial circle, Stokes' theorem converts the divergence in
Eq.~\eqref{eq:app-harrison-exact-density} into the flux
\[
  \frac{2\pi}{2\kappa}\int_{\partial\mathcal B}\dd^2x\,
                                  \sqrt{-s}\,\nu^iD_i u.
\]
Here we use the orbit domain $\mathcal B$, the induced
metric $s$, and the normal $\nu^i$ defined in
Sec.~\ref{sec:harrison-construction}. The flux cancels
against the contribution of opposite sign from the Gibbons--Hawking--York term.
The comparison is made for the same orbit domain and
the same axial period, with a smooth, axially invariant
timelike boundary. It does not cover spacelike caps or joint terms.

On the axis $V=0$, the metric $\gamma=V\bar g$ degenerates even though
the four-dimensional geometry is regular. The local treatment
first employs a tube of radius $\rho=\epsilon$. The fluxes of
the individual divergences cannot be declared to vanish:
in a flat seed with $V=\rho^2$, the inner tube satisfies
$\sqrt{-s}\,\nu^iD_i\ln V=-2$ per unit $\dd t\,\dd z$.
The cancellations must be carried out before removing the regulator.
Nor is the Gibbons--Hawking--York term of the artificial tube
part of the action of the original manifold with smooth axis.

When comparing a regular seed in smooth axial gauge with its
transform, $\Lambda=1+\Order(\rho^2)$: the flux of
$D\ln\Lambda$ and the difference between the tube terms are
$\Order(\epsilon^2)$ and disappear. This is the local argument
used to extend the comparison of actions to the axis in
the solutions of this work, without separately discarding
finite contributions of the reduction.

\subsection{Neutral seeds, normalization, and axial regularity}
\label{app:harrison-neutral-axis}

To fix the correspondence with the notation of the seeds,
we take $\bar g=\hat g^\perp$, $V=G$, $\psi=0$, and $b=1/\ell$.
Here the primes denote the magnetized physical fields:
\begin{equation}
  \Lambda=1+\frac G{\ell^2},\qquad
  g'=\Lambda^2\hat g^\perp+\frac G{\Lambda^2}\dd\varphi^2,
  \qquad A'=\frac G{\ell\Lambda}\dd\varphi.
  \label{eq:app-harrison-neutral-specialization}
\end{equation}
The seeds~\eqref{eq:harrison-Minkowski-seed} and
\eqref{eq:harrison-Schwarzschild-seed} produce the geometries
\eqref{eq:harrison-Melvin-spherical} and
\eqref{eq:schmel-sph-metric}, respectively. The function
$f=1-2m/r$ corresponds to the one that Ortaggio denotes by $f^2$.
In cylindrical coordinates of the flat seed, $G=\rho^2$ and
$A'_\varphi=b\rho^2+\Order(\rho^4)$; hence the field on
the axis is $B_0=2b=2/\ell$, with the magnetic orientation adopted.

The global extension also requires fixing the gauge and the period.
For a regular axis with transverse proper distance $\rho$,
$V=\rho^2+\Order(\rho^4)$ and a smooth potential of the form
\eqref{eq:app-harrison-ansatz} satisfies $\psi=\Order(\rho^2)$
in a regular trivialization. Then
$\Lambda=1+\Order(\rho^2)$ and the period $2\pi$ is preserved.
If one uses instead a patch with $\psi\to\psi_0$ constant,
\begin{equation}
  \Lambda_0=(1+b\psi_0)^2,\qquad
  \psi'_0=\frac{\psi_0}{1+b\psi_0},\qquad
  \Delta\varphi_{\mathrm{reg}}'=2\pi\Lambda_0^2,
  \label{eq:app-harrison-axis-period}
\end{equation}
provided that $1+b\psi_0\ne0$. The last result follows from
comparing the circumferential radius $\rho/\Lambda_0$ with the
proper radius $\Lambda_0\rho$. A constant term
$\psi_0\dd\varphi$ is not a smooth form on a disk that
includes the axis; its removal, when appropriate, must be described
with gauge patches. Moreover, shifting $\psi$ by a constant
before applying the same $H_b$ changes $\Lambda$ and can change
the resulting geometry. In Melvin and Schwarzschild--Melvin we use
the regular potential of Eq.~\eqref{eq:app-harrison-neutral-specialization},
which vanishes at both poles and avoids that ambiguity.

\subsection{Alternative check of the intrinsic identity}
\label{app:harrison-boundary-identity}

\paragraph{General axial data.}
We check the intrinsic identity through the direct
transformation of the curvature and of the induced field. This derivation
does not use the field equations and complements the intrinsic
reduction of Sec.~\ref{sec:harrison-construction}.
Let
\begin{equation}
\begin{aligned}
  h_E&=q_{AB}\dd x^A\dd x^B+V\dd\varphi^2,
  &A|_\Sigma&=\psi\,\dd\varphi,\\
  h'_E&=\Lambda^2q_{AB}\dd x^A\dd x^B
                           +\Lambda^{-2}V\dd\varphi^2,
  &A'|_\Sigma&=\psi'\,\dd\varphi,
\end{aligned}
  \label{eq:app-harrison-general-boundary}
\end{equation}
with $q$ positive definite, $V>0$, $A,B=1,2$, and $\Lambda,\psi'$ given by
Eqs.~\eqref{eq:app-harrison-complex-map}--\eqref{eq:app-harrison-general-map}.
We write $D_q$ for the connection of $q$ and contract
the scalar products between gradients with $q$. We define
$u=\ln(V/V_*)$, $\omega=\ln\Lambda$,
$\mathcal F=\dd(A|_\Sigma)$, and
$\mathcal F'=\dd(A'|_\Sigma)$; each electromagnetic invariant
is contracted with its respective three-dimensional metric. The curvature
formulas~\eqref{eq:harrison-warped-curvature} and
\eqref{eq:harrison-base-conformal} give
\begin{equation}
  \Lambda^2\cR[h'_E]-\cR[h_E]
       =2D_q\omega\cdot D_q u-2(D_q\omega)^2.
  \label{eq:app-harrison-general-curvature}
\end{equation}
On the other hand, the isometry~\eqref{eq:app-harrison-isometry}
implies, using $u'=u-2\omega$,
\begin{equation}
  \frac{2}{V'}(D_q\psi')^2-
  \frac{2}{V}(D_q\psi)^2
      =2D_q u\cdot D_q\omega-2(D_q\omega)^2.
  \label{eq:app-harrison-general-Maxwell}
\end{equation}
The left-hand side is $\Lambda^2\mathcal F'^2-\mathcal F^2$,
since $\mathcal F'^2=2(D_q\psi')^2/V$ and
$\mathcal F^2=2(D_q\psi)^2/V$. Subtracting, one obtains
\begin{equation}
  \boxed{\ \cR[h'_E]-\mathcal F'^2
        =\Lambda^{-2}\bigl(\cR[h_E]-\mathcal F^2\bigr)\ }.
  \label{eq:app-harrison-general-intrinsic-identity}
\end{equation}
Since $\sqrt{h'_E}=\Lambda\sqrt{h_E}$, in the domain
$X=\cR[h_E]-\mathcal F^2>0$ it follows that
\begin{equation}
  \sqrt{h'_E}\sqrt{2\bigl(\cR[h'_E]-\mathcal F'^2\bigr)}
    =\sqrt{h_E}\sqrt{2\bigl(\cR[h_E]-\mathcal F^2\bigr)}.
  \label{eq:app-harrison-general-density-identity}
\end{equation}
This is a pointwise identity, without integration by parts, for
axial data related by the map. The neutral seed
$\psi=0$ reduces the right-hand side to
$\sqrt{\hat h_E}\sqrt{2\hat{\cR}}$ and reproduces
Eq.~\eqref{eq:harrison-counterterm-density}. The origin of the
mixed counterterm in the transport of the geometric functional
is thus identified explicitly. The proof also shows that the sign
of $X$ is preserved as long as $\Lambda>0$.

\paragraph{Check for a neutral seed.}
We take the data~\eqref{eq:harrison-seed-boundary} and the map
\eqref{eq:harrison-boundary-map}. From here on, the hat
denotes the seed and the unprimed symbols the magnetized
fields, as in Eq.~\eqref{eq:harrison-curvature-identity};
in particular, $\hat{\cR}=\cR[\hat h_E]$. Let $\hat D_q$
be the connection of $\hat q$ and $\omega=\ln\Lambda$. The same
geometric formulas give
\begin{equation}
  \Lambda^2\cR[h_E]=\hat{\cR}
       +2\hat D_q\omega\cdot\hat D_q\ln(G/V_*)-2(\hat D_q\omega)^2.
  \label{eq:harrison-curvature-transformation}
\end{equation}
If $Q_G=\hat q^{AB}\partial_AG\partial_BG$, then
\begin{equation}
\begin{aligned}
  \partial_A\omega&=\frac{\partial_AG}{\ell^2\Lambda},
  &\partial_A A_\varphi&=\frac{\partial_AG}{\ell\Lambda^2},\\
  2\hat D_q\omega\cdot\hat D_q\ln(G/V_*)-2(\hat D_q\omega)^2
    &=\frac{2Q_G}{\ell^2\Lambda G}-\frac{2Q_G}{\ell^4\Lambda^2}
     =\frac{2Q_G}{\ell^2\Lambda^2G},\\
  \mathcal F^2
    &=\frac{2\hat q^{AB}}G\partial_A A_\varphi\partial_B A_\varphi
     =\frac{2Q_G}{\ell^2\Lambda^4G}.
\end{aligned}
  \label{eq:harrison-curvature-Maxwell-cancellation}
\end{equation}
In the second line we used $\Lambda-1=G/\ell^2$; in the third,
the conformal factors of $h_E^{AB}$ and $h_E^{\varphi\varphi}$
cancel. Substituting into Eq.~\eqref{eq:harrison-curvature-transformation}
confirms Eq.~\eqref{eq:harrison-curvature-identity}.

\paragraph{Scope of the identity.}
The equality of densities does not identify Dirichlet problems
with the same data: the Harrison transformation acts simultaneously on the
boundary metric and potential. Nor does it allow
the independent variations of the Brown--York functional to be restricted to
the Harrison orbit. For such variations one must use the
intrinsic extension~\eqref{eq:harrison-counterterm-transfer}
and the domain of differentiability
\eqref{eq:melvin-sph-E-variation-domain}. At $X=0$ the
equality of values may persist, but the general derivative
of the square root is no longer guaranteed.

\subsection{Seeds with a nonvanishing potential}
\label{app:harrison-charged-seeds}

The derivation of Eq.~\eqref{eq:app-harrison-general-intrinsic-identity}
does not assume $\psi=0$. We illustrate it on the boundaries
$r=\mathrm{const.}$ with two seeds whose induced potential does not
vanish: a seed that is itself magnetized, for which the axis remains
regular, and a magnetically charged seed, for which the pointwise
identity holds but the potential cannot vanish at both poles. In this
subsection the hat denotes the seed and the prime its image under
$H_b$. In both examples $\cR$ and $\mathcal F^2$ have been computed
directly from the metrics and potentials given, without using
Eqs.~\eqref{eq:harrison-warped-curvature}--\eqref{eq:harrison-intrinsic-reduction}.

\paragraph{Magnetized seed.}
We take as seed the Euclidean boundary data of Melvin ($f=1$) or of
Schwarzschild--Melvin ($f=1-2m/r$) with parameter $b_1=1/\ell_1$,
Eqs.~\eqref{eq:melvin-sph-induced-normal} and
\eqref{eq:schmel-sph-boundary-geometry}:
\begin{equation}
\begin{aligned}
\dd\hat s^2_{\Sigma,E}
&=\Lambda_1^2\left(f\,\dd\tau^2+r^2\dd\theta^2\right)
+\frac{r^2\sin^2\theta}{\Lambda_1^2}\,\dd\varphi^2,\\
\hat\psi&=\frac{b_1r^2\sin^2\theta}{\Lambda_1},
\qquad
\Lambda_1=1+b_1^2r^2\sin^2\theta .
\end{aligned}
\label{eq:app-charged-seed-melvin}
\end{equation}
Its induced field does not vanish:
\begin{equation}
\hat{\mathcal F}^2=\frac{8b_1^2\cos^2\theta}{\Lambda_1^4},
\qquad
\hat X=\hat{\cR}-\hat{\mathcal F}^2=\frac{2}{r^2\Lambda_1^2},
\label{eq:app-charged-seed-melvin-invariants}
\end{equation}
in agreement with Eq.~\eqref{eq:melvin-sph-intrinsic-invariants}. With
$\hat V=r^2\sin^2\theta/\Lambda_1^2$, the map $H_{b_2}$ gives
\begin{equation}
\begin{aligned}
\Lambda&=(1+b_2\hat\psi)^2+b_2^2\hat V=\frac{\Lambda_{12}}{\Lambda_1},
\qquad
\psi'=\frac{(b_1+b_2)r^2\sin^2\theta}{\Lambda_{12}},\\
\Lambda_{12}&=1+(b_1+b_2)^2r^2\sin^2\theta,
\end{aligned}
\label{eq:app-charged-seed-melvin-map}
\end{equation}
and $\dd s'^2_{\Sigma,E}=\Lambda_{12}^2(f\,\dd\tau^2+r^2\dd\theta^2)
+\Lambda_{12}^{-2}r^2\sin^2\theta\,\dd\varphi^2$. The image is therefore
the same family with parameter $b_1+b_2$, as required by
$H_{b_2}\circ H_{b_1}=H_{b_1+b_2}$
(Appendix~\ref{app:harrison-general-map}). Its invariant is
$X'=2/(r^2\Lambda_{12}^2)=\Lambda^{-2}\hat X$, and
\begin{equation}
\sqrt{h'_E}\sqrt{2X'}=\sqrt{\hat h_E}\sqrt{2\hat X}
=2r\sqrt f\,\sin\theta .
\label{eq:app-charged-seed-melvin-density}
\end{equation}

This example has three consequences. First, the identity is verified
with $\hat{\mathcal F}^2\neq0$ and $\mathcal F'^2\neq0$, so that the
electromagnetic term of the counterterm is tested on both sides and not
only generated from a neutral seed. Second, the seed functional is
$\sqrt{2(\hat{\cR}-\hat{\mathcal F}^2)}$ and the transported functional
is the same: the counterterm~\eqref{eq:melvin-sph-counterterm} is form
invariant along the Harrison orbit. Transporting $\sqrt{2\hat{\cR}}$
from Minkowski or Schwarzschild in one step, with $b=b_1+b_2$, or in
two steps, $b_1$ and then $b_2$, yields the same functional and the
same value $\beta r\sqrt f$. The construction therefore does not depend
on which member of the orbit is taken as seed, and the argument leading
to Eq.~\eqref{eq:harrison-action-invariance} applies equally to any two
members, both renormalized with Eq.~\eqref{eq:melvin-sph-counterterm}.
Third, $\hat\psi=\Order(\sin^2\theta)$ at both poles and
$\Lambda=1+b_2(2b_1+b_2)r^2\sin^2\theta+\Order(\sin^4\theta)$, so that
the period $2\pi$ is preserved, as required in
Appendix~\ref{app:harrison-neutral-axis}.

\paragraph{Magnetically charged seed.}
We take the boundary data induced at $r=\mathrm{const.}$ by the
magnetically charged Reissner--Nordstr\"om solution, with
$F=P\sin\theta\,\dd\theta\wedge\dd\varphi$ and
$f=1-2m/r+P^2/r^2$ in the normalization of
Eq.~\eqref{eq:lorentzian-bulk-GH}:
\begin{equation}
\dd\hat s^2_{\Sigma,E}
=f\,\dd\tau^2+r^2\dd\theta^2+r^2\sin^2\theta\,\dd\varphi^2,
\qquad
\hat\psi=\psi_0-P\cos\theta,
\label{eq:app-charged-seed-monopole}
\end{equation}
where the constant $\psi_0$ labels the gauge. Then
\begin{equation}
\hat{\cR}=\frac2{r^2},\qquad
\hat{\mathcal F}^2=\frac{2P^2}{r^4},\qquad
\hat X=\frac{2(r^2-P^2)}{r^4},
\label{eq:app-charged-seed-monopole-invariants}
\end{equation}
which is positive for $r>|P|$. The map $H_b$ has
\begin{equation}
\Lambda=\bigl(1+b\psi_0-bP\cos\theta\bigr)^2+b^2r^2\sin^2\theta,
\label{eq:app-charged-seed-monopole-Lambda}
\end{equation}
and the direct computation of $\cR[h'_E]$ and $\mathcal F'^2$ for the
image data~\eqref{eq:app-harrison-general-boundary} confirms
$X'=\Lambda^{-2}\hat X$ for every $\psi_0$. The transported density and
its integral are
\begin{equation}
\begin{aligned}
\sqrt{h'_E}\sqrt{2X'}&=\sqrt{\hat h_E}\sqrt{2\hat X}
=2\sqrt f\,\sqrt{r^2-P^2}\,\sin\theta,\\
\frac{\beta(2\pi)}{\kappa}\int_0^\pi\dd\theta\,
2\sqrt f\sqrt{r^2-P^2}\,\sin\theta
&=\beta\sqrt f\sqrt{r^2-P^2}.
\end{aligned}
\label{eq:app-charged-seed-monopole-density}
\end{equation}
This is the value of the seed functional
$\sqrt{2(\hat{\cR}-\hat{\mathcal F}^2)}$, which differs from
$\sqrt{2\hat{\cR}}$ through the field of the monopole. The density
depends only on $\hat{\mathcal F}$ and is independent of $\psi_0$,
whereas $\Lambda$, and therefore the image, depend on $\psi_0$.

The global use of this image is restricted by the axis. The difference
$\hat\psi_S-\hat\psi_N=2P$ between the values at the south and north
poles is gauge invariant, since $2\pi(\hat\psi_S-\hat\psi_N)$ is the
flux $4\pi P$ through the sphere. For $P\neq0$, $\hat\psi$ cannot
vanish at both poles, and the regular trivialization of
Appendix~\ref{app:harrison-neutral-axis} does not exist. Assuming
$1+b\hat\psi_N\neq0$ and $1+b\hat\psi_S\neq0$,
\begin{equation}
\Lambda_N=(1+b\psi_0-bP)^2,\qquad
\Lambda_S=(1+b\psi_0+bP)^2,\qquad
\Lambda_S-\Lambda_N=4bP(1+b\psi_0).
\label{eq:app-charged-seed-monopole-poles}
\end{equation}
By Eq.~\eqref{eq:app-harrison-axis-period}, the regular periods at the
two poles, $2\pi\Lambda_N^2$ and $2\pi\Lambda_S^2$, differ for
$bP\neq0$ unless $1+b\psi_0=0$. Outside that gauge, the image has a
conical singularity on at least one half-axis for any choice of the
period of $\varphi$. The pointwise identity holds away from the poles
in every gauge; what fails is the global regularity of the image. This
is the precise sense in which the global use of seeds with $\psi\neq0$
requires the analysis of the gauge and of the axial period. We do not
analyze the global structure of this image, including the gauge
$1+b\psi_0=0$.

\section{Exact integrals and expansions of the Euclidean action}
\label{app:euclidean-action-integrals}

We present the separate evaluations of the bulk and
Gibbons--Hawking--York terms that support
Sec.~\ref{sec:euclidean-action}. We keep the conventions
of Sec.~\ref{sec:melvin-universe}, the outward normal,
and the complete sphere. The radial expansions are performed
on the integrated expressions, not on an angular
approximation that excludes the poles.

\subsection{Melvin}

The regulated region is that of Eq.~\eqref{eq:melvin-sph-boundary},
with $0\leq\bar r\leq r$, $\ell>0$, and arbitrary period $\beta$.
We start from the action~\eqref{eq:melvin-sph-E-action}.
The trace of the Einstein--Maxwell equations gives $R=0$, whereas
\begin{equation}
  F^2=\frac{8}{\ell^2\bar\Lambda^4},\qquad
  \sqrt{g_E}=\bar\Lambda^2\bar r^2\sin\theta.
  \label{eq:melvin-sph-bulk-invariants}
\end{equation}
After integrating over $\tau$ and $\varphi$, we obtain
\begin{equation}
  I^E_{\mathrm{bulk}}(r)
  =\frac{\beta}{\ell^2}\int_0^r\dd\bar r\,\bar r^2
      \int_0^\pi\dd\theta\,
       \frac{\sin\theta}{(1+\bar r^2\sin^2\theta/\ell^2)^2}.
  \label{eq:melvin-sph-bulk-integral}
\end{equation}
The required angular integrals can be obtained with
$u=\cos\theta$. For $x\geq0$,
\begin{align}
  \int_0^\pi\frac{\sin\theta\,\dd\theta}{1+x^2\sin^2\theta}
  &=\frac{2\operatorname{arsinh}x}{x\sqrt{1+x^2}},
       \nonumber\\
  \int_0^\pi\frac{\sin\theta\,\dd\theta}{(1+x^2\sin^2\theta)^2}
  &=\frac{1}{1+x^2}
    +\frac{\operatorname{arsinh}x}{x(1+x^2)^{3/2}},
  \label{eq:melvin-sph-angular-integrals}
\end{align}
with value $2$ by continuity at $x=0$ in both cases.
The radial integration gives the exact expression
\begin{equation}
  I^E_{\mathrm{bulk}}(r)
  =\beta\left[r-
       \frac{\ell^2}{\sqrt{r^2+\ell^2}}
         \operatorname{arsinh}\!\left(\frac r\ell\right)\right].
  \label{eq:melvin-sph-bulk-exact}
\end{equation}

The induced metric and the outward normal are those of
Eq.~\eqref{eq:melvin-sph-induced-normal}; the trace of the extrinsic
curvature and its density are given in Eq.~\eqref{eq:melvin-sph-K}.
Using the first integral of Eq.~\eqref{eq:melvin-sph-angular-integrals},
\begin{equation}
\begin{aligned}
  I^E_{\mathrm{GH}}(r)
   &=-\frac{\beta r}{2}\int_0^\pi\dd\theta\,
            \sin\theta\left(2-\frac1\Lambda\right)\\
   &=-2\beta r
      +\frac{\beta\ell^2}{\sqrt{r^2+\ell^2}}
            \operatorname{arsinh}\!\left(\frac r\ell\right).
\end{aligned}
  \label{eq:melvin-sph-GH-exact}
\end{equation}
The exact sum of these contributions is
Eq.~\eqref{eq:melvin-sph-bare-exact}, presented in
Sec.~\ref{sec:euclidean-action}.

To compare with the asymptotic expansion, let
$L(r)=\ln(2r/\ell)$. At fixed $\ell$ and $\beta$,
\begin{align}
  I^E_{\mathrm{bulk}}(r)
   &=\beta r-\frac{\beta\ell^2}{r}L(r)
     +\Order\!\left(\frac{\beta\ell^4[1+L(r)]}{r^3}\right),
       \nonumber\\
  I^E_{\mathrm{GH}}(r)
   &=-2\beta r+\frac{\beta\ell^2}{r}L(r)
     +\Order\!\left(\frac{\beta\ell^4[1+L(r)]}{r^3}\right).
  \label{eq:melvin-sph-bulk-GH-asymptotics}
\end{align}
These expansions are taken after integrating. The approximation
$\Lambda\sim r^2\sin^2\theta/\ell^2$ is not uniform at the poles:
in particular, $\sin\theta/\Lambda^2\sim
\ell^4/(r^4\sin^3\theta)$ at fixed angle cannot be integrated over the whole of
$[0,\pi]$. This behavior indicates the relevance of the regions of
angular width $\Order(\ell/r)$ around the axes for the
leading term of the bulk integral.

\subsection{Schwarzschild--Melvin}

\paragraph{Regularity of the Euclidean geometry.}
For $m>0$, we regulate the exterior region with $2m\leq\bar r\leq r$,
$r>2m$, and the smooth outer boundary
$\Sigma\simeq S^1_\beta\times S^2$. In the regular Euclidean solution,
the horizon is the origin of the $(\bar r,\tau)$ plane, not an inner
boundary; regularity fixes $\beta=8\pi m$.
The Euclidean metric is
\begin{equation}
\dd s_E^2=\bar\Lambda^2\left(f(\bar r)\dd \tau^2+
\frac{\dd\bar r^2}{f(\bar r)}+\bar r^2\dd\theta^2\right)
+\frac{\bar r^2\sin^2\theta}{\bar\Lambda^2}\dd\varphi^2,
\qquad \bar\Lambda=1+\frac{\bar r^2\sin^2\theta}{\ell^2}.
\label{eq:schmel-sph-E-metric}
\end{equation}
Indeed, if $\rho^2=8m(\bar r-2m)$ and
$\Lambda_H=1+4m^2\sin^2\theta/\ell^2$, at leading order the metric
in the plane normal to the horizon is
\begin{equation}
  \dd s^2_{E,\perp}\simeq\Lambda_H^2
     \left(\dd\rho^2+\frac{\rho^2}{16m^2}\dd\tau^2\right).
  \label{eq:schmel-sph-bolt-regularity}
\end{equation}
The angular dependence of $\Lambda_H$ multiplies both
terms equally and does not change the required period. The horizon is a
two-dimensional fixed surface of the Euclidean circle; we do not add
a Gibbons--Hawking--York term there as if it were an inner
boundary~\cite{GibbonsHawking,HawkingHorowitz}.

\paragraph{Evaluation of the bulk term by means of the Maxwell field.}
The bulk contribution is

\begin{equation}
I^E_{\mathrm{bulk}}
=-\frac{1}{2\kappa}\int_{\cM}\dd^4x\,\sqrt{g_E}\,(R-F^2) .
\end{equation}

Evaluating on the solution, $R=0$ and
\begin{equation}
  F^2=\frac{8\bigl[f(\bar r)\sin^2\theta+\cos^2\theta\bigr]}
              {\ell^2\bar\Lambda^4},\qquad
  \sqrt{g_E}=\bar\Lambda^2\bar r^2\sin\theta,
  \qquad \bar\Lambda=1+\frac{\bar r^2\sin^2\theta}{\ell^2}.
  \label{eq:schmel-sph-bulk-invariants}
\end{equation}
Defining
\begin{equation}
  H(r)=\frac{\ell^2}{\sqrt{r^2+\ell^2}}
                   \operatorname{arsinh}\!\left(\frac r\ell\right),
  \label{eq:schmel-sph-H}
\end{equation}
we can evaluate the bulk term without expanding the integrand at the poles.
The Maxwell equation and $F=\dd A$ imply
\begin{equation}
  F^2=2\nabla_\mu(A_\nu F^{\mu\nu}),\qquad
  I^E_{\mathrm{bulk}}
    =\frac1\kappa\int_{\partial\cM}\dd^3x\,
          \sqrt h\,A_\nu n_\mu F^{\mu\nu}.
  \label{eq:schmel-sph-Maxwell-divergence}
\end{equation}
This is an identity on the solution, not a Legendre term
added to the action. On the outer cutoff,
\begin{equation}
  n_\mu F^{\mu\varphi}=\frac{2\sqrt f}{\ell r\Lambda},
  \qquad
  \sqrt{h_E}\,A_\varphi n_\mu F^{\mu\varphi}
       =\frac{2fr^3\sin^3\theta}{\ell^2\Lambda}.
  \label{eq:schmel-sph-Maxwell-flux}
\end{equation}
Using
$\int_0^\pi\sin\theta\,\Lambda^{-1}\dd\theta=2H(r)/r$
and $(r^2/\ell^2)\sin^2\theta/\Lambda=1-1/\Lambda$, one obtains
the exact expression and its expansion at fixed $m$ and $\ell$:

\begin{equation}
\begin{aligned}
I^E_{\mathrm{bulk}}(r)
&=\beta f(r)\bigl[r-H(r)\bigr]\\
&=\beta r-2\beta m-\frac{\beta\ell^2\ln(2r/\ell)}{r}
+\frac{2\beta m\ell^2\ln(2r/\ell)}{r^2}
+\Order\!\left(\frac{\beta\ell^4[1+\ln(r/\ell)]}{r^3}\right).
\end{aligned}
\label{eq:schmel-sph-bulk-exact}
\end{equation}

\paragraph{Gibbons--Hawking--York term.}
The induced metric, $\sqrt{h_E}$, the outward normal, and the trace
of the extrinsic curvature are the boundary data presented
in Eq.~\eqref{eq:schmel-sph-boundary-geometry} and in
Sec.~\ref{sec:melvin-universe}.

The Gibbons--Hawking--York term,
\begin{equation}
I^E_{\mathrm{GH}}
=-\frac{1}{\kappa}\int_{\Sigma}\dd^3y\,\sqrt{h_E}\,K ,
\end{equation}
has the exact expression and the asymptotic behavior
\begin{equation}
\begin{aligned}
I^E_{\mathrm{GH}}(r)
&=\beta f(r)\bigl[-2r+H(r)\bigr]-\frac{\beta m}{2}\\
&=-2\beta r+\frac{7}{2}\beta m
+\frac{\beta\ell^2\ln(2r/\ell)}{r}
-\frac{2\beta m\ell^2\ln(2r/\ell)}{r^2}
+\Order\!\left(\frac{\beta\ell^4[1+\ln(r/\ell)]}{r^3}\right).
\end{aligned}
\label{eq:schmel-sph-GH-exact}
\end{equation}

The exact cancellation of the terms containing $H(r)$ gives
the sum~\eqref{eq:schmel-sph-bare-exact}, presented in
Sec.~\ref{sec:euclidean-action}.
The above expansions have been performed after integrating over
the whole sphere. Expanding $\bar\Lambda$ first at fixed angle can
miss polar regions that contribute to the integral.

The evaluation of the counterterm and the sum of the three
contributions are presented in
Sec.~\ref{sec:euclidean-action}. In particular,
Eq.~\eqref{eq:schmel-sph-counterterm-exact} provides its
exact value and its expansion, whereas
Eqs.~\eqref{eq:schmel-sph-renormalized-exact} and
\eqref{eq:schmel-sph-renormalized-infinity} distinguish
the regulated action from its radial limit.

\section{Euclidean variation and local components of the responses}
\label{app:variation-local-components}

We collect the first variation in Euclidean signature and the
local computations that support the results of
Secs.~\ref{sec:variational-principle} and
\ref{sec:brown-york-energy}. The Euclidean expressions
are written in terms of $h_E$ and $\mathcal C^E_{ab}$;
the Brown--York components are then evaluated in
Lorentzian signature, with the conventions already fixed.

\subsection{Euclidean first variation and Dirichlet data}

The functional~\eqref{eq:melvin-sph-counterterm} depends only on the
induced metric, on the tangential potential through $\mathcal F$,
and on their intrinsic derivatives. To make its good behavior precise
it is necessary to distinguish the local regularity of the functional,
stationarity with prescribed data, and the removal of the regulator.
The value of the action on a solution does not determine its
functional derivatives.

\paragraph{Prescribed data and intrinsic character.}
We work on the fixed Euclidean boundary
$\Sigma\simeq S^1_\beta\times S^2$, with $r$, $\beta$, and $\ell$ fixed,
and admit smooth variations of the bulk fields. The
Dirichlet conditions fix their complete restrictions:
\begin{equation}
  \delta h^E_{ab}\big|_\Sigma=0,\qquad
  \delta A_a\big|_\Sigma=0.
  \label{eq:melvin-sph-E-Dirichlet-data}
\end{equation}
The potential is fixed in a regular gauge; the admissible gauge
transformations preserve those data. Keeping $\ell$ fixed identifies
the external field within the exact family, but it does not replace
the functional conditions~\eqref{eq:melvin-sph-E-Dirichlet-data}
for general perturbations.
On the outer cutoff of Schwarzschild--Melvin we fix the complete
induced metric and the tangential potential, $\delta h^E_{ab}|_\Sigma=0$ and
$\delta A_a|_\Sigma=0$.

The curvature $\cR$ contains second tangential derivatives of $h_E$,
and $\mathcal F$ contains first tangential derivatives of $A_a$.
The counterterm contains neither $K_{ab}$ nor $F_{na}$; its variation does not
introduce normal derivatives of the variations that would force
new data to be fixed. The tangential derivatives are transferred to their
coefficients by intrinsic integrations by parts.
The Gibbons--Hawking--York term cancels the normal derivatives
of the metric variation of the Einstein--Hilbert term, whereas the Maxwell term
leaves a contribution proportional to $\delta A_a$.
This is the structural reason why a functional of the induced
data can be compatible with Dirichlet conditions~\cite{GibbonsHawking,JubbEtAl:2017}.

\paragraph{Regularity at the poles and absence of additional edges.}
The boundary $\Sigma$ is compact and has no edge. The local
expression~\eqref{eq:melvin-sph-polar-potential} shows the regularity
of the potential at the poles, which are not components of
$\partial\Sigma$.

\paragraph{First variation and electromagnetic response.}
In Euclidean signature we use the definitions and the domain
$X>0$ of Eq.~\eqref{eq:melvin-sph-E-variation-domain}.
The indices and the connection $D_a$ of this paragraph refer to $h_E$.
The identities
\begin{equation}
  \delta U=\Xi(\delta\cR-\delta\mathcal F^2),\qquad
  \delta\mathcal F^2
    =2\mathcal F_{ac}\mathcal F_b{}^c\delta h_E^{ab}
       +4\mathcal F^{ab}D_a\delta A_b
  \label{eq:melvin-sph-E-mixed-variation}
\end{equation}
show that the electromagnetic dependence also modifies the
response conjugate to the potential. After integrating by parts,
\begin{equation}
\begin{aligned}
  \delta I^E_{\mathrm{ct}}
   &=\frac1\kappa\int_\Sigma\dd^3x\,\sqrt{h_E}\,
       \left[\mathcal C^E_{ab}\delta h_E^{ab}
              +4D_a(\Xi\mathcal F^{ab})\delta A_b\right],\\
  \mathcal C^E_{ab}
   &=-\frac U2h^E_{ab}
      +\Xi(\mathcal R_{ab}-2\mathcal F_{ac}\mathcal F_b{}^c)
      +(h^E_{ab}D^2-D_aD_b)\Xi.
\end{aligned}
  \label{eq:melvin-sph-E-counterterm-variation}
\end{equation}
Combining this expression with the bulk and Gibbons--Hawking--York terms,
\begin{equation}
\begin{aligned}
 \left.\delta I^E_{\mathrm{ren}}\right|_{\mathrm{EOM}}
  ={}&\frac1{2\kappa}\int_\Sigma\dd^3x\,\sqrt{h_E}\,
    \left[-(K_{ab}-Kh^E_{ab})+2\mathcal C^E_{ab}\right]\delta h_E^{ab}\\
   &+\frac2\kappa\int_\Sigma\dd^3x\,\sqrt{h_E}\,
    \left[n_\mu F^{\mu a}+2D_b(\Xi\mathcal F^{ba})\right]\delta A_a.
\end{aligned}
  \label{eq:melvin-sph-E-total-variation}
\end{equation}
Here the equations of motion are imposed on the background, not on its
variations. The contribution $D_b(\Xi\mathcal F^{ba})$ arises from
an intrinsic functional: it is not a Legendre transformation of the
Maxwell term, nor does it change the ensemble to one of fixed normal flux
\cite{BradenBrownWhitingYork}. The above equation shows that
fixing $(h^E_{ab},A_a)$ suffices for stationarity at finite cutoff;
it is not necessary to fix their conjugate momenta as well:
\begin{equation}
  \left.\delta I^E_{\mathrm{ren}}\right|_{\mathrm{EOM,\,Dirichlet}}=0.
  \label{eq:melvin-sph-finite-Dirichlet}
\end{equation}
When varying, $U$ must be kept as a functional of the induced fields.
Substituting $U=2/(r\Lambda)$ beforehand would eliminate variations of
$\cR$ and $\mathcal F^2$ and, with them, the Hessian and
current terms in Eq.~\eqref{eq:melvin-sph-E-total-variation}.
That identity on a solution is used after varying.

\subsection{Local Brown--York components on Melvin}

We detail the local evaluation of the
Brown--York responses~\cite{BrownYork}. Their tensorial,
electromagnetic, and energetic results are presented in
Sec.~\ref{sec:brown-york-energy}.

\paragraph{Lorentzian action and boundary data.}
The outer boundary is $\Sigma:\bar r=r$, with fixed $r>0$ and
$\Sigma\simeq\mathbb R_t\times S^2$. Its induced metric, in coordinates
$x^a=(t,\theta,\varphi)$, is
\begin{equation}
  h_{ab}\dd x^a\dd x^b
  =-\Lambda^2\dd t^2+\Lambda^2r^2\dd\theta^2
       +\frac{r^2\sin^2\theta}{\Lambda^2}\dd\varphi^2,
  \qquad \sqrt{-h}=\Lambda r^2\sin\theta,
  \label{eq:melvin-sph-BY-induced}
\end{equation}
where $\Lambda=1+r^2\sin^2\theta/\ell^2$ and $\kappa=8\pi$.
The poles are regular points of the closed spatial section; there are no
additional faces or responses localized at corners. We consider
smooth variations, regular at the origin and on the axis, with compact
support in $t$, so that the temporal endpoints do not contribute to
the first variation. The complete action and its first variation are presented in
Sec.~\ref{sec:variational-principle}. We evaluate the responses
\eqref{eq:melvin-sph-BY-responses}, or equivalently the form
\eqref{eq:brown-york-spherical}, after varying the functional.

\paragraph{Exact evaluation on the Melvin universe.}
The outward normal is $n^\mu\partial_\mu=
\left.\bar\Lambda^{-1}\partial_{\bar r}\right|_{\bar r=r}$.
Differentiating the components of the metric before restricting to the cutoff,
\begin{equation}
\begin{aligned}
  K_{ab}&=\left.\frac1{2\bar\Lambda}
                       \partial_{\bar r}h_{ab}(\bar r,\theta)
                \right|_{\bar r=r},\\
  K_{tt}&=-\frac{2(\Lambda-1)}r,\qquad
  K_{\theta\theta}=r(3\Lambda-2),\qquad
  K_{\varphi\varphi}
     =\frac{r\sin^2\theta(2-\Lambda)}{\Lambda^4},\\
  K&=\frac{2(2\Lambda-1)}{r\Lambda^2}.
\end{aligned}
  \label{eq:melvin-sph-BY-K-components}
\end{equation}
The Lorentzian continuation does not change the intrinsic invariants
of Eq.~\eqref{eq:melvin-sph-intrinsic-invariants}. In particular,
\begin{equation}
\begin{aligned}
  U&=\frac2{r\Lambda},\qquad \Xi=\frac{r\Lambda}{2},\qquad
  \partial_\theta\Xi=\frac{r^3\sin\theta\cos\theta}{\ell^2},\\
  D^2\Xi
   &=\frac1{\Lambda r^2\sin\theta}
         \partial_\theta\!\left(
           \frac{\sin\theta}{\Lambda}\partial_\theta\Xi\right)\\
   &=\frac{r(3\cos^2\theta-1)}{\ell^2\Lambda^2}
      -\frac{2r^3\sin^2\theta\cos^2\theta}{\ell^4\Lambda^3}.
\end{aligned}
  \label{eq:melvin-sph-BY-Xi-derivatives}
\end{equation}
The connection is included in the components of the Hessian:
$D_aD_b\Xi=\partial_a\partial_b\Xi
-\Gamma^c{}_{ab}\partial_c\Xi$. Although $\partial_t\Xi=0$, in general
$D_tD_t\Xi\neq0$; for example,
\begin{equation*}
  D_tD_t\Xi=-\frac{2r^3\sin^2\theta\cos^2\theta}{\ell^4\Lambda}.
\end{equation*}
The cancellation in the component that determines the energy can
be shown explicitly. The intrinsic curvature satisfies
\begin{equation*}
  \mathcal R_{tt}
   =\frac{2[4\cos^2\theta-\Lambda(1+\cos^2\theta)]}
           {\ell^2\Lambda^2},
  \qquad
  \Xi\mathcal R_{tt}+(h_{tt}D^2-D_tD_t)\Xi=0.
\end{equation*}
Since $\mathcal F_{tc}=0$, this gives
$\mathcal C_{tt}=-Uh_{tt}/2=\Lambda/r$.
Substituting into Eq.~\eqref{eq:melvin-sph-BY-ct-coefficient}, the two
contributions $\tau^{\mathrm{grav}}_{ab}=-(K_{ab}-Kh_{ab})/\kappa$
and $\tau^{\mathrm{ct}}_{ab}=2\mathcal C_{ab}/\kappa$ are
\begin{equation}
\begin{aligned}
  \tau^{\mathrm{grav}}_{tt}&=-\frac{2\Lambda}{\kappa r},
  &\tau^{\mathrm{ct}}_{tt}&=+\frac{2\Lambda}{\kappa r},\\
  \tau^{\mathrm{grav}}_{\theta\theta}&=\frac{r\Lambda}{\kappa},
  &\tau^{\mathrm{ct}}_{\theta\theta}&=-\frac{r\Lambda}{\kappa},\\
  \tau^{\mathrm{grav}}_{\varphi\varphi}
    &=\frac{r\sin^2\theta(5\Lambda-4)}{\kappa\Lambda^4},
  &\tau^{\mathrm{ct}}_{\varphi\varphi}
    &=-\frac{r\sin^2\theta(5\Lambda-4)}{\kappa\Lambda^4}.
\end{aligned}
  \label{eq:melvin-sph-BY-component-cancellation}
\end{equation}
The resulting cancellation is presented in
Eq.~\eqref{eq:melvin-sph-BY-ren-components}, in
Sec.~\ref{sec:brown-york-energy}.

The electromagnetic response provides an additional check. The
only components that can contribute are
\begin{equation}
\begin{aligned}
  n_\mu F^{\mu\varphi}&=\frac2{\ell r\Lambda},\qquad
  \sqrt{-h}\,\Xi\mathcal F^{\theta\varphi}
                  =\frac{r\cos\theta}{\ell},\\
  D_a(\Xi\mathcal F^{a\varphi})
     &=\frac1{\sqrt{-h}}\partial_\theta
          (\sqrt{-h}\,\Xi\mathcal F^{\theta\varphi})
       =-\frac1{\ell r\Lambda}.
\end{aligned}
  \label{eq:melvin-sph-BY-Maxwell-cancellation}
\end{equation}
These contributions give the vanishing response
\eqref{eq:melvin-sph-BY-current-zero}. Its variational scope
and the energies~\eqref{eq:melvin-sph-BY-energy-zero}
are presented in Sec.~\ref{sec:brown-york-energy}.

\subsection{Local Brown--York components on Schwarzschild--Melvin}
\label{sec:schmel-spherical}

For $m>0$, $\ell>0$, and $r>2m$, we use the geometry and the
parameters defined in Sec.~\ref{sec:melvin-universe}.

\paragraph{Responses of the complete action.}
We evaluate the responses~\eqref{eq:melvin-sph-BY-responses} of the
complete action on Schwarzschild--Melvin. The coefficient
$\mathcal C_{ab}$ is that of Eq.~\eqref{eq:melvin-sph-BY-ct-coefficient},
and the metric response admits the form~\eqref{eq:brown-york-spherical}.
The variation with respect to independent sources has been carried out
before substituting the solution, as established in
Sec.~\ref{sec:variational-principle}.

In Lorentzian signature, $h_{tt}=-\Lambda^2f$ and
$n=(\sqrt f/\Lambda)\partial_{\bar r}$. The identity
$K_{ab}=\tfrac12\mathcal L_n h_{ab}$ gives the components
\begin{align}
K_{tt}
&=
-\sqrt{f}
\left(
\frac{2rf\sin^2\theta}{\ell^2}
+\frac{m\Lambda}{r^2}
\right),
\\
K_{\theta\theta}
&=
r\sqrt{f}\,(3\Lambda-2),
\\
K_{\varphi\varphi}
&=
\frac{r\sqrt{f}\sin^2\theta}{\Lambda^4}
(2-\Lambda),
\end{align}
whereas its trace is
\begin{equation}
K
=
\frac{\sqrt{f}}{\Lambda}
\left[
\frac{4}{r}
-\frac{2}{r\Lambda}
+\frac{m}{r(r-2m)}
\right].
\label{eq:schmel-sph-BY-K-trace}
\end{equation}

The nonvanishing components of the intrinsic Ricci tensor are
\begin{align}
\cR_{tt}
&=
\frac{2f}{\ell^2\Lambda^2}
\left[
2-3\sin^2\theta
+\frac{r^2}{\ell^2}\sin^2\theta
\left(\sin^2\theta-2\right)
\right],
\\
\cR_{\theta\theta}
&=
\frac{1}{\Lambda^2}
\left[
1
+\frac{2r^2}{\ell^2}
\left(3-2\sin^2\theta\right)
+\frac{r^4}{\ell^4}\sin^2\theta
\left(3\sin^2\theta-2\right)
\right],
\\
\cR_{\varphi\varphi}
&=
\frac{\sin^2\theta}{\Lambda^6}
\left[
1
+\frac{6r^2}{\ell^2}\cos^2\theta
+\frac{r^4}{\ell^4}\sin^2\theta
\left(\sin^2\theta-2\right)
\right].
\end{align}
Since $f$ is tangentially constant, $\mathcal R_{tt}$ and
$D_tD_t\Xi$ are $f$ times their values in Melvin; the spatial
components and $D^2\Xi$ do not change. In particular,
$\Xi=r\Lambda/2$ and
\begin{equation}
  \Xi\mathcal R_{tt}+(h_{tt}D^2-D_tD_t)\Xi=0,
  \qquad \mathcal C_{tt}=-\frac U2h_{tt}=\frac{f\Lambda}{r}.
  \label{eq:schmel-sph-BY-temporal-cancellation}
\end{equation}
The cancellation uses the complete Hessian; $\partial_t\Xi=0$
does not allow $D_tD_t\Xi$ to be discarded. The other components are
\begin{equation}
  \mathcal C_{\theta\theta}=-\frac{r\Lambda}{2},
  \qquad
  \mathcal C_{\varphi\varphi}
     =-\frac{r\sin^2\theta(5\Lambda-4)}{2\Lambda^4}.
  \label{eq:schmel-sph-BY-spatial-counterterm}
\end{equation}
In this way the response of the tangential field and of the
curvature is incorporated, without introducing an extrinsic background subtraction.

\paragraph{Exact coordinate components and radial expansions.}
Substituting into Eq.~\eqref{eq:melvin-sph-BY-responses}, the off-diagonal
components vanish and one obtains
\begin{equation}
\begin{aligned}
\tau^{\mathrm{ren}}_{tt}
={}&\frac{2f\Lambda}{\kappa r}\left(1-\sqrt{f}\right)
\\
={}&\frac{2m\sin^2\theta}{\kappa\ell^2}
-\frac{3m^2\sin^2\theta}{\kappa\ell^2 r}
\\
&+\frac{1}{\kappa r^2}
\left(
2m-\frac{m^3\sin^2\theta}{\ell^2}
\right)
+\Order\!\left(r^{-3}\right) .
\end{aligned}
\label{eq:schmel-sph-BY-temporal-component}
\end{equation}
The polar component is
\begin{equation}
\begin{aligned}
\tau^{\mathrm{ren}}_{\theta\theta}
={}&\frac{r\Lambda}{2\kappa\sqrt{f}}
\left(1-\sqrt{f}\right)^2
\\
={}&\frac{m^2\sin^2\theta}{2\kappa\ell^2}\,r
+\frac{m^3\sin^2\theta}{\kappa\ell^2}
\\
&+\frac{1}{\kappa r}
\left(
\frac{m^2}{2}+\frac{15m^4\sin^2\theta}{8\ell^2}
\right)+\Order(r^{-2}).
\end{aligned}
\label{eq:schmel-sph-BY-polar-component}
\end{equation}
and the azimuthal component is
\begin{equation}
\begin{aligned}
\tau^{\mathrm{ren}}_{\varphi\varphi}
={}&\frac{r\sin^2\theta}{\kappa\Lambda^4}
\left[
(5\Lambda-4)\left(\sqrt{f}-1\right)
+\frac{\Lambda(1-f)}{2\sqrt{f}}
\right]
\\
={}&-\frac{4m\ell^6}{\kappa\sin^4\theta}\,\frac{1}{r^6}+\Order(r^{-7}) .
\end{aligned}
\label{eq:schmel-sph-BY-azimuthal-component}
\end{equation}
The expansions are taken at fixed $m$ and $\ell$. The first two
admit remainders that are uniform in $\theta$, since their angular dependence
is polynomial in $\sin^2\theta$. The last one has been written for
fixed $0<\theta<\pi$: the factor $\sin^{-4}\theta$ prevents its use
at the poles. In all cases, the exact expressions are the ones
used to integrate or to examine the axis.

The physical components~\eqref{eq:schmel-sph-BY-physical-components}
and their uniform control~\eqref{eq:schmel-sph-BY-uniform-bounds}
are presented in Sec.~\ref{sec:brown-york-energy}.

\paragraph{Electromagnetic response and conservation.}
The response conjugate to $A_a$ is needed to check the
consistency of the metric components. On this background,
\begin{equation}
  n_\mu F^{\mu\varphi}=\frac{2\sqrt f}{\ell r\Lambda},
  \qquad
  D_a(\Xi\mathcal F^{a\varphi})=-\frac1{\ell r\Lambda},
  \label{eq:schmel-sph-BY-Maxwell-pieces}
\end{equation}
The resulting response~\eqref{eq:schmel-sph-BY-current}
is presented in Sec.~\ref{sec:brown-york-energy}.

We use the boundary identities~\eqref{eq:melvin-sph-BY-Ward}
of the complete action.
Their polar component provides an independent check:
\begin{equation}
  D_a\tau^{\mathrm{ren}\,a}{}_{\theta}
    =\frac{8r(1-\sqrt f)\sin\theta\cos\theta}
            {\kappa\ell^2\Lambda^3}
    =\mathcal F_{\theta\varphi}\mathcal J^\varphi_{\mathrm{ren}}.
  \label{eq:schmel-sph-BY-Ward-polar}
\end{equation}
The interpretation of this identity and the conservation in time
are presented in Sec.~\ref{sec:brown-york-energy},
together with the proper energy~\eqref{eq:schmel-sph-BY-proper-energy}
and the charge~\eqref{eq:schmel-sph-BY-charge-exact}.

\bibliography{references}

\end{document}